\documentclass[prb,amsmath,superscriptaddress,citeautoscript,twocolumn,
showkeys,showpacs,floatfix]{revtex4}
\usepackage{bm}
\usepackage{amssymb}
\usepackage{graphicx}

\begin{document}

\title{Voltage-induced singularities in transport through molecular junctions}

\author{O. Entin-Wohlman}
\email{oraentin@bgu.ac.il}

\altaffiliation{Also at Tel Aviv University, Tel Aviv 69978,
Israel}


 \affiliation{Department of Physics, Ben
Gurion University, Beer Sheva 84105, Israel}

\affiliation{Albert Einstein Minerva Center for Theoretical
Physics, Weizmann Institute of Science, Rehovot 76100, Israel}

\author{Y. Imry}

\affiliation{Department of Condensed Matter Physics,  Weizmann
Institute of Science, Rehovot 76100, Israel}

\author{A. Aharony}

\altaffiliation{Also at Tel Aviv University,
Tel Aviv 69978, Israel}

\affiliation{Department of Physics and the Ilse Katz Center for
Meso- and Nano-Scale Science and Technology, Ben Gurion
University, Beer Sheva 84105, Israel}

\date{\today}
\begin{abstract}
The inelastic scattering of electrons which carry current through
a single-molecule junction is modeled by a quantum dot, coupled to
electron reservoirs via  two  leads.  When the electron is on the
dot, it is coupled to a single harmonic oscillator of frequency
$\omega_0$.  At zero temperature, the resonance peak in the
linear-response conductance always narrows down due to the
coupling with the vibrational mode. However, this narrowing down
is given by the Franck-Condon factor only for narrow resonances.
Contrary to some claims in the literature, the linear-response
conductance does not exhibit any side-bands at zero temperature.
Small side-bands, of order $\exp [-\beta \hbar \omega_0]$, do
arise at finite temperatures.
The single-particle density of states exhibits discontinuities and
logarithmic singularities at the frequencies  corresponding to the
opening of the inelastic channels,  due to the imaginary and real
parts of the self-energy.
The same singularities also generate discontinuities and
logarithmic divergences in the differential conductance at and
around the inelastic thresholds. These discontinuities usually 
involve upwards steps, but these steps become
negative within a rather narrow range of the elastic transparency 
of the junction. This range shrinks further as the the excitation 
energy exceeds the bare resonance width.

\end{abstract}



\maketitle

\section{ Introduction}

Single-molecule junctions based on direct bonding of a small
molecule between two metallic electrodes  seem  by now rather
established experimentally.
\cite{REED,reichert,zhitenev,kubatkin,QIU,kushmerick,pasupathy,dekker,djukic}
The electronic transport through such a molecular bridge is
attracting a great deal of interest, including the invention of
ingenious experimental realizations for it (see, for example, the
recent Refs. ~\onlinecite{Tal}). Besides the possible technological
advantages of ``molecular electronics" \cite{Aviram-Ratner}, there
are many issues that make this problem of great interest from both
the basic science and the application points of view. The
possibility of directly addressing a {\em single} microscopic
quantum system with an ordinary measurement apparatus should shed
light on fundamental quantum measurement questions.

Electrons passing through the small molecule may change its
quantum state (electronic, vibrational, and in certain cases also
rotational, and even the  conformation \cite{EGGER-CONF}  of the
molecule). These may require a finite energy transfer from the
transport electron. Thus, the dynamics of the molecule may create
interesting structures in the I-V characteristics.
\cite{NDR,SWITCH} These rich characteristics, resembling the one
observed in inelastic electron tunneling spectroscopy (IETS),
\cite{KIR} depend on important experimental details such as  the
equilibration time of the vibrations compared to the typical time
between consecutive electrons passing through the junction,   or
whether the electrons can pump the molecule into higher
vibrational states. Such measurements provide a handle on studying
molecular properties and their modifications by the binding to the
electrodes. In some cases they may also help to identify the
molecule which has been bound in the bridge.

The configurational modification of the molecule by the tunneling
electron is usually described by a linear coupling  of the electron
with e.g.  the vibrational modes, while the oscillating location of
the whole molecule is modeled  by the dependence of the tunneling
matrix elements to the leads on the vibrational degrees of freedom.
\cite{MITRA} As is well-known, one may eliminate the linear
electron-phonon interaction by a canonical transformation which
dresses the tunneling matrix elements by the phonon cloud (the
Holstein polaron \cite{HOLS}). The resulting matrix elements contain
the Franck-Condon factors. These tend to block the conductance at
off-resonant situations (the Franck-Condon blockade).
\cite{AJI,FELIX} However, the top of the resonance conductance is
not reduced by these factors. \cite{MITRA} Indeed, we find  that the
coupling to the vibrational mode causes a narrowing of the resonance. This
narrowing is described by the `usual' Franck-Condon blockade only in
the limit of very narrow `bare' resonances, namely very long dwell
times of the electrons on the resonances. We find that
reducing this dwell time weakens the Franck-Condon blocking.

Transport through small molecules offers new means of studying
tunneling of electrons interacting with vibrational modes.
Theoretical studies of the coupling between molecular vibrations and
electronic states participating in the tunneling have begun with the
exact calculation of the single-electron transmission, in which
the presence of the Fermi seas representing the leads has been
essentially ignored. \cite{GLAZMAN,WINGREEN} The single-electron
transmission naturally exhibits resonances at energies corresponding
to the vibration frequencies (these are often called ``side-bands").
Such side-bands also appear in the local (on-molecule)
single-particle density of states, computed in the presence of the
leads. \cite{FLENSBERG,THOSS} There are claims in the literature
\cite{LUNDIN,BALATSKY} that these side-bands are reflected in, for
example, the gate-potential dependence of the linear-response
conductance. However, as has been  emphasized by Mitra {\it et al.}
\cite{MITRA} and discussed in detail below, such side-bands cannot
appear at zero temperature  in the linear-response regime. We shall
demonstrate their appearance, albeit weakly, at finite temperatures.

Indeed,  an electron crossing  the molecular bridge may do so
inelastically or elastically (with or without  changing the
excitation state of the molecule). In the former case the electron
will lose its phase coherence -- a problem to which we will return
in future work (see the discussion in Ref. ~\onlinecite{DORON}).
Here, we concentrate on the structure of the conductance as a
function of the bias voltage $V$ and the gate potential,
represented by  the electrochemical potential $\mu$ (when
applying the latter is feasible). Clearly, at low temperatures only
elastic processes and inelastic ones {\em exciting } the molecule
are possible. The latter can happen only if the transmitted electron
can supply the energy required for the molecular excitation.
Focussing on a molecular vibration of frequency $\omega^{}_{0}$, it
is clear, then, that it can be excited only when the bias voltage
$V$ exceeds $\hbar \omega^{}_{0}/e$, namely, beyond the
linear-response regime. This \cite{MITRA}  will be confirmed by the
detailed calculations below.

The footprints of the inelastic processes appear in the differential
conductance when plotted as a function of the bias voltage. (For
analyses of the full counting statistics of a vibrating junction,
see Refs. ~\onlinecite{FC}.)  This regime has been studied
experimentally rather intensively. Theoretically, it has been
treated by employing a variety  of methods and numerical techniques.
\cite{MAC,NITZAN,munich,CHEM,ryndyk} At low temperatures, the
inelastic channel comes in when the bias voltage exceeds $\hbar
\omega^{}_{0}/e$. This however does not necessarily imply an
increase of the total conductance,  since the
elastic conduction channel might be modified as well. Indeed,
interestingly enough, it has been observed that the
 ``step"  in the conductance at $V=\hbar \omega^{}_{0}/e$  appears
 either as a decrease or an increase in the differential conductance. \cite{zhitenev,pasupathy,dekker,djukic,Tal}
Theoretical work
addressing this issue \cite{paulson,vega} claimed that this
behavior depends in a universal manner on the bare elastic
transparency of the junction, ${\cal T}$, such that the
differential conductance steps upwards when ${\cal T}<1/2$, and
downwards when ${\cal T}>1/2$. This claim has been refuted
recently in a seminal paper by Egger and Gogolin. \cite{gogolin}
We confirm their conclusion. Moreover, we find that the
conductance steps downwards only in a narrow range of ${\cal T}$,
which becomes narrower as the ratio of the excitation energy
$\hbar\omega^{}_0$ to the bare resonance width $\Gamma^{}_0$
increases.

Another important aspect concerns the instabilities in the vibration
modes possibly induced by the current.
\cite{MITRA,JONSON,KOCH} In particular, Ref.~~\onlinecite{KOCH}
points out the inapplicability of the perturbation theory in the
electron-vibration coupling once the nonequilibrium regime is
reached. We show below that the step-like  structure in the
differential conductance at $V=\hbar \omega^{}_{0}/e$ implies
another type of breakdown of the perturbation theory.  It turns out
that the opening of the inelastic channel is inevitably accompanied
by the appearance of {\em logarithmic singularities} at the same
bias voltage. Those are forced via the Kramers-Kronig relations and
are related to the singularities found by Engelsberg and Schrieffer
\cite{SCHR} for bulk Einstein phonons. In this way we confirm the
important (and seemingly, un-noticed) result of Mitra {\it et al.}
\cite{MITRA} and Egger and Gogolin \cite{gogolin}: beside the
step-like structure, caused by the inelastic tunneling processes,
the differential conductance develops a logarithmic singularity (at
zero temperature, and to second order in the electron-vibration
coupling) as the bias voltage crosses the vibration energy.  Near
the threshold voltage $V=\hbar \omega^{}_{0}/e$,  that singularity
dominates the differential conductance.

It thus seems that there are several relevant issues in the theory
of transport through a vibrating junction which are either still
under debate or are not entirely clear. These concern the existence
of side-bands, the dependence of the conductance on the junction
transparency, the structure of the differential conductance near the
opening of the inelastic channel, and the precise effect of the
Franck-Condon factors on the resonances, including what happens when
the resonance width exceeds the vibration frequency. Below, we give
our answers to these questions and provide further physical
interpretations for them. In order not to obscure the basic physics
by lengthy computations, we restrict ourselves to the simplest
model, of a single resonance connected symmetrically to two
leads and coupled linearly to a vibration. In addition,
we apply lowest-order perturbation theory in the electron-vibration
coupling. We believe that a complete analytical discussion of the
outcome of this model will shed further light on the intriguing
non-equilibrium behavior of the vibration-induced conductance.

Section \ref{MODEL} gives the Hamiltonian, and then expresses the
current through the system in terms of the Green functions, which
contain the contributions from the coupling to the vibrational mode. The
detailed calculation of these Green functions is described in the
Appendix. Section \ref{LR} presents the results for the
conductance and for the density of states in the linear-response
regime, while Sec. \ref{DIFG} discusses  the
differential conductance at finite bias voltage (but zero
temperature). Finally, we detail our conclusions in Sec. \ref{SUM}.

\section{The model}
\label{MODEL}

We consider the differential conductance of a small system,
consisting of two leads connected together via a ``dot". The two
leads are assumed to be identical, except for being attached to
reservoirs held at possibly different chemical potentials,
$\mu_{L}\equiv \mu+eV/2$ and $\mu_{R}\equiv \mu-eV/2$. When the
electron is on the dot, it is coupled to a single harmonic
oscillator of frequency $\omega^{}_{0}$. The Hamiltonian of this
system is
\begin{align}
{\cal H}={\cal H}^{}_{\rm lead}+{\cal H}^{}_{\rm dot}+{\cal H}^{}_{\rm coup}\ .
\end{align}
The lead Hamiltonian is [using $k(p)$ for the left (right) lead,
with the same lattice constant $a=1$]
\begin{align}
{\cal H}^{}_{\rm lead}=\sum_{k}\epsilon^{}_{k}c^{\dagger}_{k}c^{}_{k}+
\sum_{p}\epsilon^{}_{p}c^{\dagger}_{p}c^{}_{p}\  ,\label{HDOT}
\end{align}
with
\begin{align}
\epsilon_{k(p)}^{}=-2J\cos k(p)\  .\label{TBE}
\end{align}
The Hamiltonian of the dot is
\begin{align}
{\cal H}^{}_{\rm
dot}=\epsilon^{}_{0}c^{\dagger}_{0}c^{}_{0}+\hbar \omega^{}_{0}\bigl (
b^{\dagger}b+\frac{1}{2}\bigr )+\gamma
(b+b^{\dagger})c^{\dagger}_{0}c^{}_{0}\ , \label{HD}
\end{align}
where  $\epsilon_{0}$ is the energy level on the dot, and $\gamma$ is the coupling energy of the electron (while it resides on the dot) with the oscillator. Below we often set $\hbar=1$.
Finally, the coupling between the dot and the leads  is described  by
\begin{align}
{\cal H}^{}_{\rm coup}=\sum_{k}V^{}_{k}(c^{\dagger}_{k}c^{}_{0}+{\rm hc})+\sum_{p}V^{}_{p}(c^{\dagger}_{p}c^{}_{0}+{\rm hc})\ ,
\end{align}
with
\begin{align}
V^{}_{k(p)}=-\sqrt{\frac{2}{N}}J^{}_{0}\sin k(p)\ .\label{VK}
\end{align}
(The wave functions on the leads are normalized assuming that each
lead consists of $N$  sites.) In Eqs. (\ref{TBE}) and (\ref{VK}),
$J$ is the overlap amplitude   along the leads and $J_{0}$ is the
overlap amplitude  between the leads and the dot (taken to be symmetric, for
simplicity), all in units of energy. The operators
$c^{\dagger}_{0}$, $c^{\dagger}_{k}$, and $c^{\dagger}_{p}$
($c^{}_{0}$, $c^{}_{k}$, and $c^{}_{p}$) create (annihilate) an
electron on the dot, on the left lead, and on the right lead,
respectively, while $b^{\dagger}$ ($b$) creates (annihilates) an
excitation of the harmonic oscillator, of frequency
$\omega^{}_{0}$. This model system  has gained much theoretical
interest before, see for example Refs. ~\onlinecite{MITRA},
~\onlinecite{AJI}, ~\onlinecite{FELIX}, ~\onlinecite{GLAZMAN},
~\onlinecite{WINGREEN}, ~\onlinecite{FLENSBERG},
~\onlinecite{LUNDIN}, ~\onlinecite{BALATSKY},
~\onlinecite{NITZAN}, ~\onlinecite{CHEM} and
~\onlinecite{gogolin}.

\subsection{The currents in the system}
\label{CURRENTS}

The currents flowing in this system can be expressed in terms of
the Keldysh Green functions: those on the dot are marked by the
subscript 00, and the mixed ones are marked by the subscripts
$k(p)0$, see Appendix \ref{GREENFK} for details. The current
entering the dot from the left lead, $I^{}_{LD}$, is
\begin{align}
I_{LD}^{}&=e\int\frac{d\omega}{2\pi}\sum_{k}V^{}_{k}[G^{<}_{k0}(\omega )
-G^{<}_{0k}(\omega )]\nonumber\\
&=ie\int\frac{d\omega}{2\pi}\Gamma^{}_{0}(\omega )\nonumber\\
&\times
\Bigl (-G^{<}_{00}(\omega)+f^{}_{L}(\omega )[G^{a}_{00}(\omega )-G^{r}_{00}(\omega )]\Bigr )\ ,\label{ILD}
\end{align}
and that from the right one, $I_{RD}$, is
\begin{align}
I_{RD}^{}&=e\int\frac{d\omega}{2\pi}\sum_{p}V^{}_{p}[G^{<}_{p0}(\omega )
-G^{<}_{0p}(\omega )]\nonumber\\
&=ie\int\frac{d\omega}{2\pi}\Gamma^{}_{0}(\omega )\nonumber\\
&\times
\Bigl (-G^{<}_{00}(\omega)+f^{}_{R}(\omega )[G^{a}_{00}(\omega )-G^{r}_{00}(\omega )]\Bigr )\ .\label{IRD}
\end{align}
In Eqs. (\ref{ILD}) and  (\ref{IRD}),
$f_{L,R}(\omega)=1/[e^{\beta(\omega-\mu^{}_{L,R})}+1]$ are the Fermi
distributions in the two reservoirs, and $\Gamma^{}_{0}$ is the
imaginary part of the self-energy $\Sigma^{}_{0}$ due to  the
coupling with the leads, Eqs. (\ref{LSE})  and (\ref{GAM0}).
Obviously, current conservation requires  $I^{}_{LD}+I^{}_{RD}$ to
vanish. Indeed, upon adding Eqs. (\ref{ILD}) and (\ref{IRD})
[and employing Eqs. (\ref{XX}) and (\ref{GAM0})] we find
that current is conserved. Hence, the net current $I$ can be obtained as
the difference between the two currents, $I^{}_{LD}$ and
$I^{}_{RD}$. This leads to\begin{align} I
=e\int\frac{d\omega}{2\pi}\Gamma^{}_{0}(\omega )[f^{}_{L}(\omega
)-f^{}_{R}(\omega )]{\rm Im}G^{a}_{00}(\omega )\ .\label{IGOGs}
\end{align}
 This well-known exact result in which the local density of states
on the dot, ${\rm Im}G^{a}_{00}(\omega )$, contains all of its
dynamics, including the coupling to the oscillator, is  similar
to the result as given, e.g., in Refs.  ~\onlinecite{MITRA},
~\onlinecite{FLENSBERG}, ~\onlinecite{LUNDIN}, ~\onlinecite{gogolin}
and ~\onlinecite{MW}. It is customary to use the expression (\ref{IGOGs}) also for
nonlinear transport. We remark that this is valid only for bias
voltages that are not too large. \cite{LAN} Equation (\ref{IGOGs})
neglects the effects of the finite field on the system, such as the
nonlinear screening, the induced changes in $\epsilon_{0}$ and
$J_{0}$, and the possibility, mentioned above, of ``pumping" the
molecule into higher states. {\em Only when all the above
finite-voltage corrections are neglected, does the Keldysh
formulation justify using this result also in the nonlinear regime.}

The coupling with the harmonic oscillator affects the dot Green
functions, $G^{a}_{00}$ and $G^{r}_{00}$.
In the absence of the coupling
to the vibrations, the `bare' dot Green function is given by
\begin{align}
{\cal G}^{r}_{00}=\frac{1}{\omega
-\epsilon^{}_{0}-\Sigma^{r}_{0}}\ .\label{CAL0}
\end{align}
Neglecting the frequency dependence of the self-energy due to the
coupling with the leads, $\Sigma^{}_{0}$, (this is the ``wide-band
approximation") and absorbing  ${\rm Re}\Sigma^{}_{0}$  into
$\epsilon^{}_{0}$, i.e., $\epsilon^{}_{0}\rightarrow
\epsilon^{}_{\rm res}=\epsilon^{}_{0}+{\rm Re}\Sigma^{}_{0}$, the
zeroth-order Green function, Eq. (\ref{CAL0}), becomes
\begin{align}
{\cal G}^{r}_{00}(\omega )=\frac{1}{\omega -\epsilon^{}_{\rm
res}+i\Gamma^{}_{0}}\ .\label{CAL1}
\end{align}
Expanding the Green function up to order $\gamma^{2}$ yields
\begin{align}
G^{\stackrel{a}{r}}_{00}
&={\cal G}^{\stackrel{a}{r}}_{00}+({\cal G}^{\stackrel{a}{r}}_{00})^{2}\Bigl (\Delta\epsilon^{}_{0}+\Sigma^{\stackrel{a}{r}}_{\rm ho}\Bigr )\\
&={\cal G}^{\stackrel{a}{r}}_{00}+({\cal
G}^{\stackrel{a}{r}}_{00})^{2}\Bigl (\Delta E \pm i{\rm Im}
\Sigma^{a}_{\rm ho}\Bigr )\ ,\label{CURES}
\end{align}
where $\Sigma^{}_{\rm ho}$ is the self-energy due to the coupling
to the oscillator, $\Delta\epsilon^{}_{0}$ is the shift of the
energy, Eqs. (\ref{ZERO}) and (\ref{SHIFT}),
\begin{align}
\Delta\epsilon^{}_{0}=-\frac{2\gamma^{2}_{}}{\omega^{}_{0}}\int\frac{d\omega}{2\pi}|{\cal G}^{r}_{00}(\omega )|^{2}\Gamma^{}_{0}(\omega )[f^{}_{L}(\omega )+f^{}_{R}(\omega )]\ ,\label{POLARON}
\end{align}
and we have defined $\Delta E=\Delta \epsilon^{}_{0}+{\rm
Re}\Sigma^{a}_{\rm ho}$.

From the expansion Eq. (\ref{CURES}) it follows
that the current can be written in the form\cite{COMMENT}
\begin{align}
I=I^{}_{0}+I^{}_{\rm co}+I^{}_{\rm inco}\ ,
\end{align}
where $I^{}_{0}$ is the current in the absence of the coupling
with the oscillator,
\begin{align}
 I^{}_{0}
=e\int\frac{d\omega}{2\pi}\Gamma^{}_{0}(\omega )[f^{}_{L}(\omega
)-f^{}_{R}(\omega )]{\rm Im}{\cal G}^{a}_{00}(\omega )\
,\label{I0}
\end{align}
$I^{}_{\rm co}$ is the current involving the (real) shift in the
resonant level (which depends on the frequency and the chemical
potentials),
\begin{align}
I^{}_{\rm co}
&=-ie\int\frac{d\omega}{4\pi}\Gamma^{}_{0}(\omega )[f^{}_{L}(\omega )-f^{}_{R}(\omega )]\nonumber\\
&\times \bigl [\bigl ({\cal G}^{a}_{00}(\omega )\bigr )^{2}_{}-\bigl ({\cal
G}^{r}_{00}(\omega )\bigr )^{2}_{}\bigr ]\Delta E\ ,\label{Ico}
\end{align}
and $I^{}_{\rm inco}$ is the current involving the imaginary part
of $\Sigma^{}_{\rm ho}$,
\begin{align}
I^{}_{\rm inco}
&=e\int\frac{d\omega}{4\pi}\Gamma^{}_{0}(\omega )[f^{}_{L}(\omega )-f^{}_{R}(\omega )]\nonumber\\
&\times \bigl [\bigl ({\cal G}^{a}_{00}(\omega )\bigr )^{2}_{}+\bigl ({\cal
G}^{r}_{00}(\omega )\bigr )^{2}_{}\bigr ]{\rm Im}\Sigma^{a}_{\rm ho}(\omega )\
.\label{INco}
\end{align}

Below we mainly consider  zero temperature. (The effects of a
finite temperature on the linear-response conductance are
considered in Sec. \ref{TEMP}.) Furthermore, we ignore the
explicit dependence of $\Gamma^{}_0$ on $\omega$. At zero
temperature, the zeroth-order current is
\begin{align}
I^{}_{0}&=\frac{e}{2\pi}\int_{\mu^{}_{R}}^{\mu_{L}^{}}d\omega\frac{\Gamma^{2}_{0}}{\omega ^{2}+\Gamma^{2}_{0}}\nonumber\\
&=\frac{e\Gamma_{0}^{}}{2\pi}\Bigl ({\rm
arctan}\frac{\mu^{}_{L}}{\Gamma^{}_{0}}-{\rm
arctan}\frac{\mu^{}_{R}}{\Gamma^{}_{0}}\Bigr )\ ,\label{I01}
\end{align}
the current due to the effective shift in the resonance energy,
Eq. (\ref{Ico}), is [see Eq. (\ref{CAL1})]
\begin{align}
I^{}_{\rm co}
=&\frac{e\Gamma^{2}_{0}}{\pi}\int_{\mu^{}_{R}}^{\mu_{L}^{}}d\omega\frac{\omega}{(\omega^{2}+\Gamma^{2}_{0})^{2}}
 \Delta E(\omega ,\mu^{}_{L},\mu^{}_{R})\nonumber\\
 \equiv&\frac{e\Gamma^{2}_{0}}{\pi}\int_{\mu^{}_{R}}^{\mu_{L}^{}}d\omega\frac{\omega}{(\omega^{2}+\Gamma^{2}_{0})^{2}}
 \bigl (\Delta \epsilon^{}_{0}(\mu^{}_{L},\mu^{}_{R})\nonumber\\
 &\ \ \ \ \ \ \ \ \ \ \ +{\rm
Re}\Sigma^{a}_{\rm ho}(\omega ,\mu^{}_{L},\mu^{}_{R})\bigr )\ ,\label{ICOT}
\end{align}
and the current due to the imaginary part of the self-energy, Eq.
(\ref{INco}), is
\begin{align}
I^{}_{\rm inco}&=\frac{e\Gamma^{}_{0}}{2\pi}
\int_{\mu^{}_{R}}^{\mu^{}_{L}}d\omega
\frac{\omega^{2}-\Gamma^{2}_{0}}{(\omega^{2}+\Gamma^{2}_{0})^{2}}{\rm
Im}\Sigma^{a}_{\rm ho}(\omega ,\mu^{}_{L},\mu^{}_{R})\ .\label{INCOT1}
\end{align}

\subsection{The zero temperature Green functions  and self-energies}

The detailed calculations of the contributions to the Green
functions due to the coupling with the oscillator are given in the
Appendix.
At zero
temperature, Eq. (\ref{POLARON}) becomes
\begin{align}
\Delta\epsilon^{}_{0}&=-\frac{\gamma^{2}\Gamma^{}_{0}}{\pi\omega^{}_{0}}\Bigl
(\int^{\mu^{}_{L}}+\int^{\mu_{R}^{}}\Bigr )\frac{d\omega}{(\omega
-\epsilon^{}_{\rm res})^{2}+\Gamma^{2}_{0}}\ .\label{NOGOG}
\end{align}
(It seems that this shift was overlooked in Ref.~
\onlinecite{gogolin}.) In computing the explicit expressions of
the currents and the
conductances [see Secs. \ref{LR} and \ref{DIFG}], it is expedient
\cite{gogolin} to measure the frequencies  $\omega$ and $\omega
'$, as well as the chemical potentials $\mu_{L(R)}$    from
$\epsilon^{}_{\rm res}$. We then find
\begin{align}
\Delta\epsilon^{}_{0}&=-\frac{\gamma^{2}}{\omega^{}_{0}}-\frac{\gamma^{2}}{\pi\omega^{}_{0}}\sum_{\alpha
=L,R}{\rm arctan}\frac{\mu^{}_{\alpha}}{\Gamma^{}_{0}}\
.\label{NOGOG1}
\end{align}
With the same notations, Eq. (\ref{SIGHORsof})  gives
\begin{align}
&{\rm Im}\Sigma^{a}_{\rm ho}(\omega
)=\gamma^{2}_{}\Gamma^{}_{0}\Bigl (\frac{[f^{}_{L}(\omega
+\omega^{}_{0})+f^{}_{R}(\omega +\omega^{}_{0})]/2}{(\omega
+\omega^{}_{0})^{2}+\Gamma^{2}_{0}}
\nonumber\\
&+ \frac{1-[f^{}_{L}(\omega -\omega^{}_{0})+f^{}_{R}(\omega
-\omega^{}_{0})]/2}{(\omega -\omega^{}_{0} )^{2}+\Gamma^{2}_{0}}\Bigr )\nonumber\\
&=\frac{\gamma^{2}\Gamma^{}_{0}}{2}\sum_{\alpha =L,R}\Bigl (\frac{\Theta (\mu^{}_{\alpha}
-\omega^{}_{0}-\omega )}{(\omega
+\omega^{}_{0})^{2}+\Gamma^{2}_{0}} +\frac{\Theta (\omega -\mu^{}_{\alpha}
-\omega^{}_{0})}{(\omega -\omega^{}_{0})^{2}+\Gamma^{2}_{0}}\Bigr
)\ ,\label{Imsigho}
\end{align}
which reproduces the result of
Ref.~
\onlinecite{gogolin}.  Clearly, ${\rm Im}\Sigma^{a}_{\rm
ho}(\omega )=0$ unless $\omega <\mu_{L}-\omega^{}_{0}$ and/or
$\omega
>\mu_{R}+\omega^{}_{0}$.
Since this self-energy is required within an integral for which
$\mu_{R}\leq\omega\leq\mu_{L}$ [see Eq. (\ref{INCOT1})], its
contribution to the current appears only when the bias voltage exceeds
$\hbar \omega^{}_{0}/e$.  Indeed, substituting Eq. (\ref{Imsigho})
in Eq. (\ref{INCOT1}) yields
\begin{align}
I^{}_{\rm inco}&=\frac{e\gamma^{2}\Gamma^{2}_{0}}{4\pi}\Theta (\mu^{}_{L}-\mu^{}_{R}-\hbar\omega^{}_{0})\nonumber\\
&\times\Bigl (\int_{\mu^{}_{R}}^{\mu^{}_{L}-\omega^{}_{0}}d\omega \frac{\omega^{2}-\Gamma^{2}_{0}}{(\omega^{2}+\Gamma^{2}_{0})^{2}}\frac{1}{(\omega +\omega^{}_{0})^{2}+\Gamma^{2}_{0}}\nonumber\\
&+\int_{\mu^{}_{R}+\omega^{}_{0}}^{\mu^{}_{L}}d\omega
\frac{\omega^{2}-\Gamma^{2}_{0}}{(\omega^{2}+\Gamma^{2}_{0})^{2}}\frac{1}{(\omega
-\omega^{}_{0})^{2}+\Gamma^{2}_{0}}\Bigr )\ .\label{INCOT}
\end{align}
For
$\omega^{}_{0} \gg \Gamma^{}_{0}$ the integrand in Eq. (\ref{INCOT})
contains the two Lorentzians shifted from the usual resonance by
$\pm \omega^{}_{0}$. The $\Theta$-function factor determines how
much these Lorentzians contribute to the current. This reinforces the notion
that $I^{}_{\rm inco}$ is the current due to inelastic processes
where a vibration quantum is given to or taken from the oscillator
by the transmitted electron.
As we discuss below, finite temperatures result in small
contributions to the current $I^{}_{\rm inco}$ even in the
linear-response limit of zero bias voltage.

In a similar way, the real part of the self-energy is found from  Eq. (\ref{SIGHORsof})
\begin{widetext}
\begin{align}
{\rm Re}\Sigma^{a}_{\rm ho}(\omega )&=\gamma^{2}_{}\Gamma^{}_{0}\int\frac{d\omega '}{\pi}\frac{1}{\omega '^{2}_{}
+\Gamma^{2}_{0}}\Biggl  (\frac{ 1-[f^{}_{L}(\omega ')+f^{}_{R}(\omega ')]/2}{\omega
-\omega^{}_{0}-\omega '}
+\frac{ [f^{}_{L}(\omega ')+f^{}_{R}(\omega ')]/2}{\omega +\omega^{}_{0}-\omega '}\Biggr )\nonumber\\
&=\frac{\gamma^{2}}{2}\Bigl [\Bigl (\frac{\omega-
\omega^{}_{0}}{(\omega-\omega^{}_{0})^{2}+\Gamma^{2}_{0}}+\frac{\omega+\omega^{}_{0}}{(\omega
+\omega^{}_{0})^{2}+\Gamma^{2}_{0}}\Bigr )
+\frac{1}{\pi}\Bigl
(\frac{\omega+\omega^{}_{0}}{(\omega+\omega^{}_{0})^{2}+\Gamma^{2}_{0}}-\frac{\omega
-\omega^{}_{0}}{(\omega-\omega^{}_{0})^{2}+\Gamma^{2}_{0}}\Bigr )\sum_{\alpha =L,R}{\rm arctan}\frac{\mu^{}_{\alpha}}{\Gamma^{}_{0}}\nonumber\\
&+\frac{\Gamma^{}_0}{2\pi}\Bigl (
\frac{1}{(\omega+\omega^{}_{0})^{2}+\Gamma^{2}_{0}}\sum_{\alpha
=L,R}\ln\frac{\mu^{2}_{\alpha}+\Gamma^{2}_{0}}{(\omega
-\mu^{}_{\alpha}+\omega^{}_{0})^{2}}-
\frac{1}{(\omega-\omega^{}_{0})^{2}+\Gamma^{2}_{0}}\sum_{\alpha
=L,R}\ln\frac{\mu^{2}_{\alpha}+\Gamma^{2}_{0}}{(\omega
-\mu^{}_{\alpha}-\omega^{}_{0})^{2}}\Bigr )\Bigr ]\ ,\label{DELE1}
\end{align}
again reproducing the result of Ref.~ \onlinecite{gogolin}. [A
simple interpretation of Eqs.  (\ref{Imsigho}) and (\ref{DELE1})  is
given at the end of the Appendix, following
Eq. (\ref{SIGHORsof})].  As mentioned, it is convenient to
introduce [see Eq. (\ref{CURES})] the total energy shift which
depends on the frequency and on the chemical potentials $\mu_{L}$
and $\mu_{R}$,
\begin{align}
\Delta E(\omega ,\mu^{}_{L},\mu^{}_{R})&=\Delta\epsilon^{}_{0}(\mu^{}_{L},\mu^{}_{R})+{\rm Re}\Sigma^{a}_{\rm ho}(\omega ,\mu^{}_{L},\mu^{}_{R})\nonumber\\
&=\Delta\mu (\omega ,
\mu^{}_{L},\mu^{}_{R})+\frac{\gamma^{2}\Gamma^{}_{0}}{4\pi}\sum_{\alpha
=L,R}\Bigl (\frac{\ln [(\omega -\mu^{}_{\alpha
}-\omega^{}_{0})^{2}/\omega^{2}_{0} ]}{(\omega
-\omega^{}_{0})^{2}+\Gamma^{2}_{0}}-\frac{\ln [(\omega
-\mu^{}_{\alpha }+\omega^{}_{0})^{2}/\omega^{2}_{0} ]}{(\omega
+\omega^{}_{0})^{2}+\Gamma^{2}_{0}}\Bigr )\ ,\label{DELE}
\end{align}
where
\begin{align}
\Delta\mu^{}_{ }(x,\mu^{}_{L},\mu^{}_{R})=&\frac{\gamma^{2}}{[(x -\omega^{}_{0})^{2}+\Gamma^{2}_{0}][(x +\omega^{}_{0})^{2}+\Gamma^{2}_{0}]}\Bigl (x (x^{2}+\Gamma^{2}_{0}-\omega^{2}_{0})\nonumber\\
&-\frac{x\omega^{}_{0}\Gamma^{}_{0}}{\pi}\sum_{\alpha =L,R}
\ln\Bigl
[\frac{\mu^{2}_{\alpha}+\Gamma^{2}_{0}}{\omega^{2}_{0}}\Bigr
]-[(x^{2}+\Gamma^{2}_{0})^{2}-\omega^{2}_{0}(x^{2}-\Gamma^{2}_{0})]\frac{1}
{\pi\omega^{}_{0}}\sum_{\alpha =L,R}{\rm
arctan}\frac{\mu^{}_{\alpha}}{\Gamma^{}_{0}}\Bigr )\
.\label{DELTAMU}
\end{align}
\end{widetext}
The  factor $-\gamma^{2}/\omega^{}_{0}$, i.e., the polaron binding
energy [see Eq. (\ref{NOGOG1})], is independent of the frequency
and of the chemical potentials. Therefore, we may safely absorb it
in $\epsilon^{}_{\rm res}$ and omit it from Eq. (\ref{DELE}).
Inspection of Eq. (\ref{DELE}) reveals that $\Delta E$ diverges
logarithmically at $\omega =\mu^{}_{\alpha}\pm\omega^{}_{0}$. This
divergence \cite{gogolin} is dictated by the Kramers-Kronig
relations once the imaginary part of the self-energy attains a
discontinuity  [see the discussion following Eq. (\ref{Imsigho})].
The logarithmic divergence affects the conductance only in the
nonlinear regime, and disappears in the linear-response one.
However, the density of states is affected by these singularities
even in the linear-response regime, see Sec. \ref{LR}. In any
case, the logarithmic divergence implies that one should not
ignore the frequency dependence of
 $\Delta E$ and absorb this energy in $\epsilon^{}_{\rm res}$, as is sometimes done.

\section{The linear-response regime}

\label{LR}

\subsection{Zero temperature conductance}

 In the linear-response regime  the bias voltage energy $eV$ is the
smallest energy, and at zero temperature the energy shift and the
self-energy are required only at  $\omega =\mu_{L}=\mu_{R} \equiv
\mu$, where $\mu$ is the common Fermi energy of the leads
(measured from the resonance energy $\epsilon^{}_{\rm res}$). Then,
$I^{}_{\rm inco}=0$, and $I^{}_{\rm co}$ [Eq.
(\ref{ICOT})] requires the energy shift $\Delta
E(\mu,\mu,\mu)=\Delta\mu(\mu,\mu,\mu)$ [see Eq. (\ref{DELE})],
which is a smooth function of $\mu$. Thus, the only contribution
to the conductance from the coupling to the oscillator is
\begin{align}
\frac{2\pi}{e^{2}}G^{}_{\rm co}\Big |^{}_{\rm
lin}=\frac{2\mu\Gamma^{2}_{0}}{(\mu^{2}+\Gamma^{2}_{0})^{2}}\Delta
E(\mu ,\mu ,\mu )\ . \label{Gcolin}
\end{align} In the
linear-response regime the zeroth-order conductance, $G_{0}$,  is
\begin{align}
\frac{2\pi}{e^{2}}G^{}_{0}\Big |^{}_{\rm lin}=\frac{\Gamma^{2}_{0}}{\mu^{2}+\Gamma^{2}_{0}}\ ,
\end{align}
and we may combine $G_{0}$ and $G_{\rm co}$ to obtain
\begin{align}
\frac{2\pi}{e^{2}}G\Big |^{}_{\rm lin}=\frac{\Gamma^{2}_{0}}{[\mu -\Delta E(\mu ,\mu ,\mu )]^{2}+\Gamma^{2}_{0}}\ .\label{GLIN}
\end{align}
Obviously, this expression is valid up to second-order in the
coupling with the oscillator.

\begin{figure}[h]
\includegraphics[width=7cm]{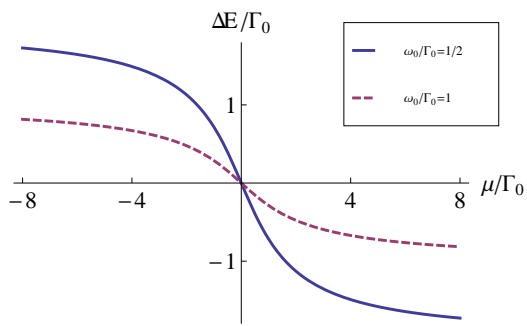}
\caption{The function $\Delta E(\mu ,\mu ,\mu )$, Eq.
(\ref{DELE}), for two representative values of the oscillator
frequency,  $\omega^{}_{0}=0.5\Gamma^{}_{0}$ (solid line), and  $\omega^{}_{0}=\Gamma^{}_{0}$ (dashed line). Here
$\gamma =\Gamma^{}_{0}$.} \label{FIGDELE}
\end{figure}

\begin{figure}[h]
\vspace{3mm}
\includegraphics[width=7cm]{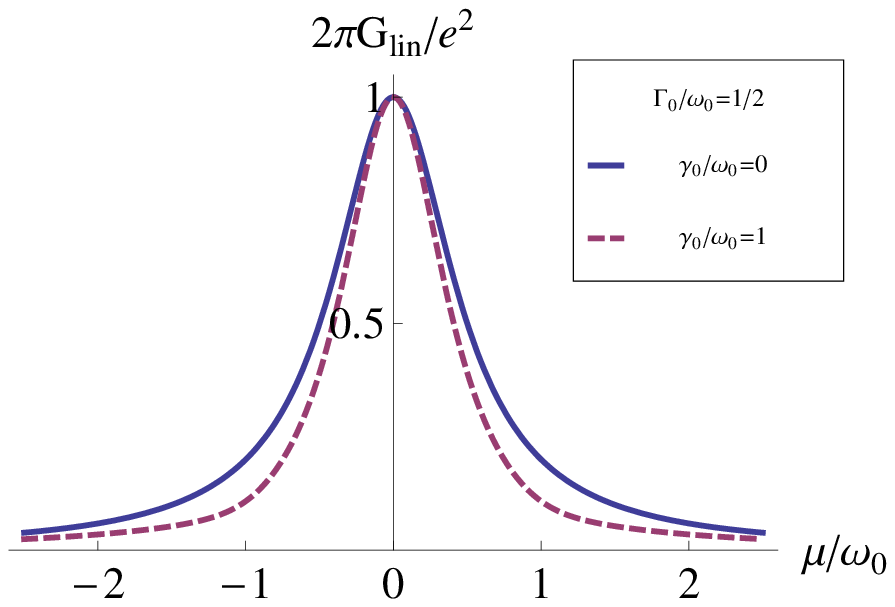}\\
\vspace{3mm}
\includegraphics[width=7cm]{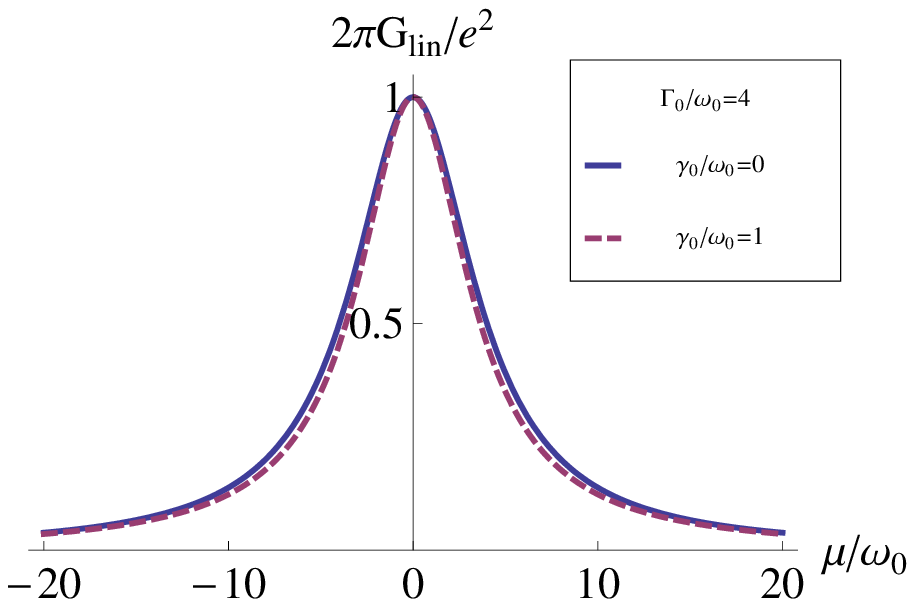}
\caption{The dimensionless linear-response conductance,  Eq.
(\ref{GLIN}), for two representative values of the ratio
$\Gamma^{}_{0}/\omega^{}_{0}$, 0.5 (top panel) and 4 (bottom
panel). It can be seen that the conductance (dashed line) is {\em always }
smaller than that obtained in the absence of the coupling with the
oscillator (depicted by the solid line).
\cite{COMMENTMITRA}} \label{RESCON}
\end{figure}

The function $\Delta E(\mu ,\mu ,\mu )$ is an odd function of
$\mu$, and its sign is opposite to that of $\mu$, see Fig.
\ref{FIGDELE}. Therefore, in the linear-response regime, the
effect of the coupling with the oscillator is just to reduce the
conductance, {\em except} at resonance. This can be understood
qualitatively as due to the fact that the couplings to the two
leads ($J_{0}$ in our model) are renormalized downwards, to ${\cal
O}(\gamma^{2})$, due to the {\em same} Franck-Condon-type factor.
Therefore, the width of the resonance decreases, but its height,
determined by the ratio of the two couplings, is
unchanged\cite{MITRA}. The location of the resonance is {\em not}
shifted, since $\mu-\Delta E(\mu ,\mu ,\mu)=0$ only at $\mu =0$,
i.e., $-\Delta E$ always moves away from the ``bare" resonance
energy. This behavior is exemplified in Fig. \ref{RESCON}, which
shows the effect of coupling to the oscillator on the linear-response
conductance, for two values of the `bare' width,
$\Gamma^{}_0/\omega^{}_0=1/2$ (top) and $4$ (bottom).

Clearly, the relative narrowing of the resonance due to the
vibrational mode decreases when the ratio $\Gamma^{}_{0}/\omega^{}_{0}$
increases. Quantitatively, the `renormalized' width of the
resonance, $\Gamma$, is given by the solution of the equation
$\Gamma-\Delta E(\Gamma,\Gamma,\Gamma)=\Gamma^{}_0$. As can be
seen from Fig. \ref{FIGDELE}, $\Delta E(\mu,\mu,\mu)$ becomes
saturated at large $\mu$;  the solution for $\Gamma$
increases towards $\Gamma^{}_0$ as $\Gamma^{}_0/\omega^{}_0$
increases. For a qualitative understanding of this effect, we note
that at zero temperature, the oscillator is in the ground state.
Then, when the electron moves from, say, the left lead to the
virtual state on the dot, the term $\gamma (b+b^{\dagger})
c^{\dagger}_{0}c^{}_{0}$ in the dot Hamiltonian, Eq. (\ref{HD}),
shifts the center of the oscillator motion by the order of
$\gamma/\omega^{}_{0}$ (in units of the oscillator's zero-point
displacement). However, this shift is fully realized only when the
dwell-time of the electron on the dot, $\Gamma^{-1}_{0}$, is
longer than the response-time of the oscillator, governed by
$\omega^{}_{0}$, i.e. $\Gamma^{}_0/\omega^{}_0 \ll 1$. In this
limit, the coupling matrix element $J_{0}$ will be reduced by the
overlap integral between the shifted and the un-shifted oscillator
wave functions, which is of order
$\exp[-(\gamma/\omega^{}_{0})^{2}/2]$, resulting in a relative
narrowing of the resonance.\cite{FELIX}

Indeed, at small $\gamma/\omega^{}_0$ (so that the perturbative
expansion is valid) and for small $\Gamma^{}_0/\omega^{}_0$ we find
that $\Gamma/\Gamma^{}_0$ approaches the Franck-Condon factor
$\exp[-(\gamma/\omega^{}_{0})^{2}]$ (which becomes
$1-(\gamma/\omega^{}_{0})^{2}$ in our order $\gamma^{2}$
approximation). This can be seen directly from Eqs. (\ref{DELE1}):
when $\omega^{}_0\gg \{|\mu|,~\Gamma^{}_0\}$,
 ${\rm Re}\Sigma^{a}_{\rm ho}(\omega )$ is dominated by the first term in
the large brackets on the second line, which is independent of the
Fermi functions. Therefore, in this limit $\Delta E(\mu,\mu,\mu)$ does not depend
on the many-body effects contained in these Fermi functions, and the
simple single particle Franck-Condon result is reproduced. Indeed,
in this limit one has $\Delta E(\mu,\mu,\mu)=\Delta\mu(\mu,\mu,\mu)
\approx -\mu\gamma^2/\omega_0^2+{\cal
O}(\mu\Gamma^{}_0\gamma^2/\omega_0^3)$, and therefore
$\Gamma\approx\Gamma^{}_0(1-\gamma^2/\omega_0^2)$. In contrast, when
$\Gamma^{}_{0}\gg\omega^{}_{0}$ (bottom panel of Fig. \ref{RESCON})
the electron leaves the dot before the oscillator has responded to
its presence, and the Franck-Condon blockade effect is much
weakened. In our calculation, part of this blocking involves the Fermi
functions on the leads [all the terms except the first in Eq. (\ref{DELE1})].
This dependence on the chemical potentials in the leads reflects the
many-body effects on the leads, which seem to weaken the Franck-Condon blockade.
So far, we have
discussed the Franck-Condon narrowing only for zero temperature and zero bias voltage.
However, the modified narrower shape of the resonances will also
affect integrals over energy, causing apparently similar effects  at finite
temperatures and at a  finite bias voltage.


Another remarkable aspect is that  at zero temperature the
linear-response conductance exhibits no side-bands as a  function
of the gate voltage (modeled here by the common $\mu$ measured
from $\epsilon^{}_{\rm res}$), when $\mu$ crosses the oscillator
frequency. This has been emphasized  in Ref. ~\onlinecite{MITRA},
contrary to certain findings in the literature (see for example
Refs. ~\onlinecite{FLENSBERG} and ~ \onlinecite{BALATSKY}). Finite
temperatures may generate small satellites, as  discussed in Sec.
\ref{TEMP}. The absence of the  side-bands in the linear-response
conductance at the oscillator frequency,  as the gate voltage  is
swept, may appear at first sight somewhat surprising.  However, it
is their appearance at zero bias voltage and zero temperature which is in
fact un-physical. A structure in the linear-response conductance
at $\mu =\pm\omega^{}_{0}$ will mean that after passing, the
electron leaves the dot in an excited state, even at zero
temperature. As the electron begins and ends at almost the same
energy, energy conservation does not allow it to excite the
oscillator.

\subsection{The zero-temperature density of states}

The situation is very different when one looks at  the local single-particle density
of states on the dot, $N(\omega )$,
given by
\begin{align}
N(\omega )=-\frac{1}{\pi}{\rm Im}G^{}_{00}(\omega )\ .
\end{align}
This density of states is accessible, in principle, via local
STM $I-V$ measurements. We note that $N(\omega )$  is the nontrivial
part of the integrand in the basic Eq. (\ref{IGOGs}) for the current.
Here we actually calculate it only at equilibrium, which is
appropriate for the linear transport regime. For
$\mu_{L}=\mu_{R}=\mu$, this quantity becomes
\begin{align}
N(\omega )=\frac{1}{\pi}\frac{\Gamma^{}_{0}+{\rm Im}\Sigma^{a}_{\rm ho}(\omega )}{[\omega -\Delta E(\omega ,\mu ,\mu)]^{2}+[\Gamma^{}_{0}+{\rm Im}\Sigma^{a}_{\rm ho}(\omega )]^{2}}\ ,\label{LOC}
\end{align}
where ${\rm Im}\Sigma^{a}_{\rm ho}(\omega )$ is given by Eq.
(\ref{Imsigho}), and  $\Delta E(\omega ,\mu,\mu) $ is given by Eqs.
(\ref{DELE1}), (\ref{DELE}), and (\ref{DELTAMU}). Inspection of
those  expressions reveals that when $\mu=0$, i.e., the common
chemical potential of the leads is aligned with the resonance
level on the dot, the density of states Eq. (\ref{LOC}) is even in
the frequency, while at off resonance (where $\mu\neq 0$ in our
notations) it is not. In the first case, there will be equal
weights for a hole (an electron) excitation corresponding to an
excited oscillator and an electron (a hole). In the second, those
weights are not equal. In particular, when $\mu>0$, i.e., the
common chemical potential of the leads is above the level on the
dot, and there is more weight to the hole formation.


\begin{figure}[h]
\includegraphics[width=7cm]{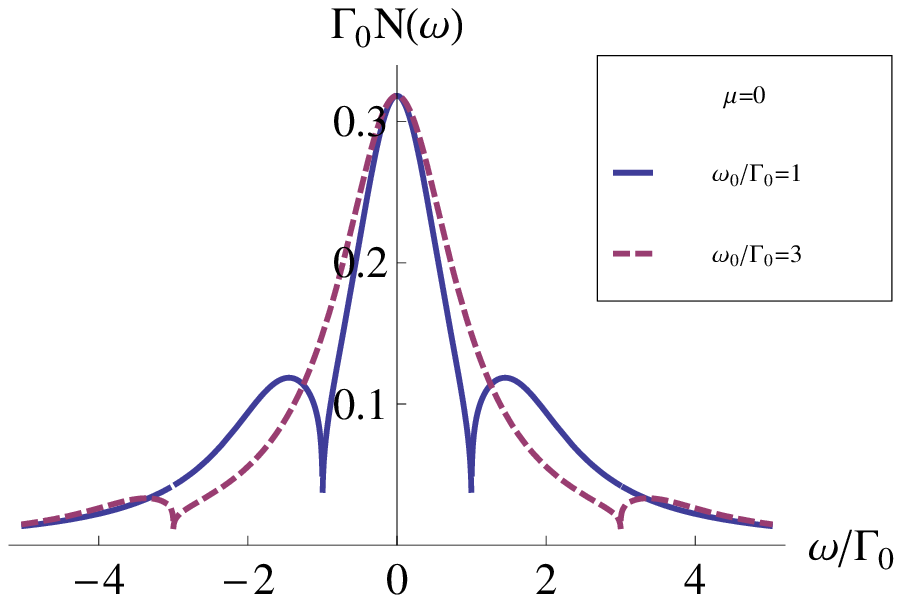} \includegraphics[width=7cm]{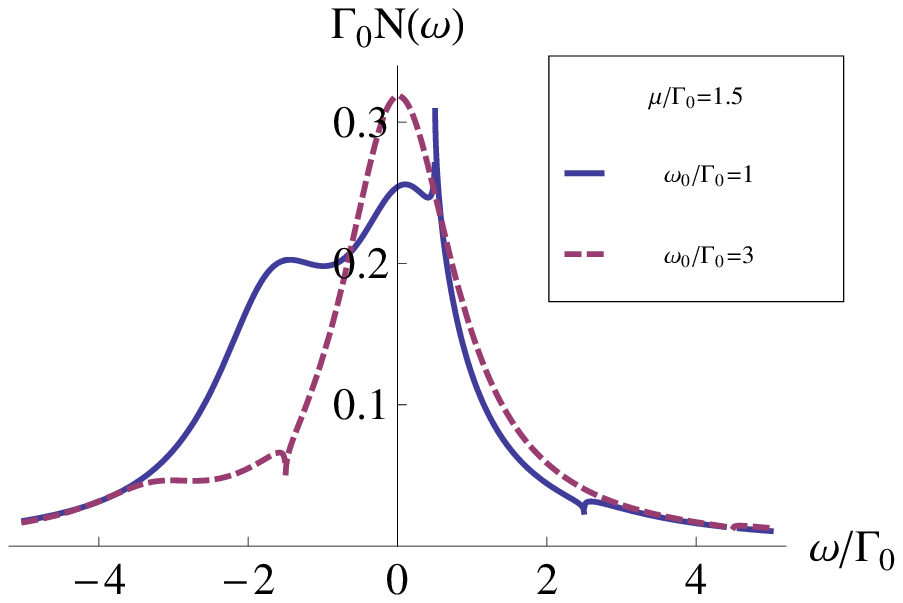}
\caption{The local density of states on the dot,  Eq. (\ref{LOC}),
as a function of the energy $\omega$ for  $\mu
=0$ (top panel) and for $\mu =1.5\Gamma_{0} $ (bottom panel). The solid lines correspond to $\omega_{0}=\Gamma_{0}$, while $\omega_{0}=3\Gamma_{0}$ for the dashed lines  (here , $\gamma
=\Gamma^{}_{0}$). All the graphs should go to zero at $\omega=\pm\omega^{}_0$.} \label{DENSTA}
\end{figure}


The local density of states $N(\omega )$ is plotted in Fig.
\ref{DENSTA} at resonance,  $\mu=0$ (upper panel), and
off-resonance, $\mu=1.5\Gamma^{}_0$ (lower panel). Both figures
show structures around $\omega=\mu\pm \omega^{}_0$, which become
smaller as $\omega^{}_0$ increases. Each of these structures
contains two ingredients: first, $\Delta E$ diverges logarithmically
 near $\omega=\mu\pm \omega^{}_0$, resulting in the vanishing of $N(\omega)$ at these frequencies.
 Since these singularities are very narrow, the plots miss showing the actual vanishing of $N(\omega)$ at these points.
 The situation is somewhat more complicated for $\mu>0$ and $\omega=\mu-\omega^{}_0$ (corresponding to the vicinity
 of $\omega_0/\Gamma^{}_0=0.5$ for the full line in the lower panel of Fig. \ref{DENSTA}). In that case, the energy
 difference $\omega-\Delta E(\omega,\mu,\mu)$ changes sign as one approaches the singular point, and therefore $N(\omega)$ first increases
 and only then decreases quickly to zero at $\omega=\mu-\omega^{}_0$. The plot picks up the initial increase, and misses the very narrow dip.

The second effect arises from ${\rm Im}\Sigma^{a}_{\rm ho}(\omega
)$, which modifies the width of the original resonance and
creates the inelastic resonances. As can be seen from Eq.
(\ref{Imsigho}), this term contains contributions from two
`resonances', at $\omega=\pm\omega^{}_0$. However, the left (right)
hand side resonance is included only for $\omega<\mu-\omega^{}_0$
($\omega>\mu+\omega^{}_0$). For $|\mu|<\omega^{}_0$, this causes a
discontinuous  increase in $N(\omega)$ for $\omega$ below (above)
$\mu-\omega^{}_0$ ($\mu+\omega^{}_0$). The deep dips at
$\omega=\mu\pm \omega^{}_0$ and the increased density of states
beyond these energies create peaks in $N(\omega)$ at
$\omega>\mu+\omega^{}_0$ and at $\omega<\mu-\omega^{}_0$, which can
be identified as the side-bands (see top panel in Fig.
\ref{DENSTA}). For $\mu>\omega^{}_0$, the behavior for
$\omega\lesssim\mu-\omega^{}_0$ is more complex, but the general
features remain the same (lower panel in the figure). Note that our
calculation shows only two such ``side-bands", since we work to
second order in the coupling with the oscillator. Note also that the
side-bands would not be as clear had we absorbed $\Delta E$ as a
`constant' in $\epsilon^{}_{\rm res}$.

\subsection{Finite temperatures}

\label{TEMP}

\begin{figure}[ht]
\vspace{4mm}
\includegraphics[width=8cm]{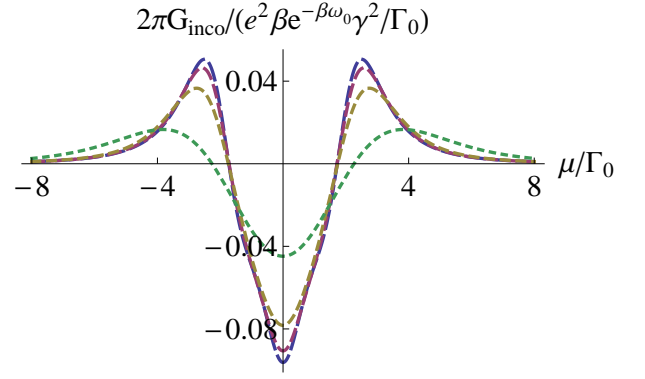}
\caption{The linear-response conductance resulting from $I^{}_{\rm
inco}$, for $\beta\Gamma^{}_0=1,~3,~6$ and 12 (increasing dash
sizes). Here $\omega^{}_{0}=2\Gamma^{}_{0}$, and all energies are
in units of $\Gamma^{}_{0}$.} \label{GincoT}
\end{figure}

At finite temperatures, the linear-response conductance  is given by [see Eq. (\ref{IGOGs})]
\begin{align}
\frac{2\pi}{e^{2}}G\big |^{}_{\rm lin}=\Gamma^{}_{0}\int d\omega \Bigl (\beta\frac{e^{\beta (\omega -\mu )}}{(e^{\beta (\omega -\mu)}+1)^{2}}\Bigr ){\rm Im }G^{a}_{00}(\omega )\ .
\end{align}
Using the expansion of Eq.  (\ref{CURES}), and the expressions in
Eq. (\ref{SIGHORsof}), and substituting $\mu_{L}=\mu_{R}=\mu$ for
linear-response, yields
\begin{align}
\Delta E(\omega )&=\gamma^{2}\int\frac{d\omega '}{\pi}\frac{\Gamma^{}_{0}}{\omega '^{2}+\Gamma^{2}_{0}}\frac{\omega -\omega '}{(\omega -\omega ')^{2}-\omega ^{2}_{0}}\nonumber\\
&\times\Bigl ({\rm coth}\frac{\beta\omega^{}_{0}}{2}+\frac{\omega -\omega '}{\omega^{}_{0}}{\rm tanh}\frac{\beta (\omega '-\mu)}{2}\Bigr )\ ,
\end{align}
and
\begin{align}
&{\rm Im}\Sigma^{a}_{\rm ho}(\omega  )=\frac{\gamma^{2}_{}}{2}\sum_{s=\pm}\frac{\Gamma^{}_{0}}{(\omega-s\omega^{}_{0})^{2}+\Gamma^{2}_{0}}\nonumber\\
&\times\Bigl ({\rm coth}\frac{\beta\omega^{}_{0}}{2}+s{\rm
tanh}\frac{\beta (\omega -s\omega^{}_{0}-\mu)}{2}\Bigr )\
.\label{finT}
\end{align}
Figure \ref{GincoT} portrays the contribution of $I^{}_{\rm
inco}$, Eq. (\ref{INco}), to the linear-response conductance. We
plot only this contribution, which arises from the inelastic
processes,  in order to exhibit the channel-opening due to the
finite temperature. The contribution of $I^{}_{\rm co}$ is smooth,
so it does not have drastic effects at finite temperature. Scaling
$I^{}_{\rm inco}$ by $\beta\Gamma^{}_0 \exp[-\beta\omega_0]$, it
is seen that the curves plotted for various temperatures approach
an asymptotic limiting form for large $\beta\Gamma^{}_0$,
exhibiting a reduction of the conductance near $\mu=0$ and peaks
slightly above (below) $\mu=\omega_0$ ($\mu=-\omega_0$). This
structure could have been described as having side-bands; however,
the peaks decay exponentially (as $\exp [-\beta\omega_0]$) at low
temperatures. This factor arises directly from the low temperature
behavior of the large brackets in Eq. (\ref{finT}), and is also
understandable intuitively: the side-bands can contribute only if
excitations by the oscillator energy $\hbar\omega^{}_0$ are
allowed. Those appear with the Boltzmann factor $\exp
[-\beta\omega_0]$. As the temperature increases, the structure
portrayed in Fig. \ref{GincoT}  broadens and gradually becomes
smeared.

\section{The zero temperature differential conductance}

\label{DIFG}

The differential conductance is the derivative of the current with
respect to the bias voltage, $V=(\mu^{}_L-\mu^{}_R)/e$ [however, for
finite bias voltage, note the discussion following Eq. (\ref{IGOGs})].
Differentiating Eq. (\ref{I01}) with respect to $V$, the
zeroth-order conductance is
\begin{align}
\frac{2\pi}{e^{2}}G^{}_{0}=&\frac{1}{2}\Bigl (\frac{\Gamma^{2}_{0}}{\mu^{2}_{L}+\Gamma^{2}_{0}}
+\frac{\Gamma^{2}_{0}}{\mu^{2}_{R}+\Gamma^{2}_{0}}\Bigr )\ .\label{GZEROR}
\end{align}
Similarly, differentiating  Eq. (\ref{ICOT}) gives
\begin{align}
\frac{2\pi}{e^{2}}G^{}_{\rm co}=
\frac{2\pi}{e^{2}}G^{(1)}_{\rm co}
+
\frac{2\pi}{e^{2}}G^{(2)}_{\rm co}\ ,
\end{align}
where
\begin{align}
\frac{2\pi}{e^{2}}G^{(1)}_{\rm co}=\sum_{\alpha =L,R}\frac{\Gamma^{2}_{0}\mu^{}_{\alpha}}{(\mu^{2}_{\alpha}+\Gamma^{2}_{0})^{2}}\Delta E(\omega =\mu^{}_{\alpha},\mu^{}_{L},\mu^{}_{R})\ ,\label{GCO1}
\end{align}
and
\begin{align}
\frac{2\pi}{e^{2}}G^{(2)}_{\rm
co}=2\Gamma^{2}_{0}\int_{\mu^{}_{R}}^{\mu_{L}}d\omega\frac{\omega}{(\omega^{2}+\Gamma^{2}_{0})_{}^{2}}
\frac{d\Delta E(\omega )}{d(\mu^{}_{L}-\mu^{}_{R})}\
.\label{DIFICO}
\end{align}
Using Eq. (\ref{DELE}), we have
\begin{align}
\frac{d\Delta E(\omega )}{d(\mu^{}_{L}-\mu^{}_{R})}&=\frac{\gamma^{2}\Gamma^{}_{0}}{2\pi}\Bigl (\frac{1}{\mu^{2}_{R}+\Gamma^{2}_{0}}\frac{(\omega -\mu^{}_{R})^{2}/\omega ^{}_{0}}{(\omega -\mu^{}_{R})^{2}-\omega^{2}_{0}}\nonumber\\
&
-\frac{1}{\mu^{2}_{L}+\Gamma^{2}_{0}}\frac{(\omega -\mu^{}_{L})^{2}/\omega ^{}_{0}}{(\omega -\mu^{}_{L})^{2}-\omega^{2}_{0}}\Bigr )\ .\label{DELV}
\end{align}
A rather lengthy computation yields
\begin{widetext}
\begin{align}
\frac{2\pi}{e^{2}}G^{}_{\rm co}&=\sum_{\alpha =L,R}\frac{\mu^{}_{\alpha}\Gamma^{2}_{0}\Delta \mu (\mu^{}_{\alpha},\mu^{}_{L},\mu^{}_{R})}{(\mu^{2}_{\alpha}+\Gamma^{2}_{0})^{2}}
+\frac{\gamma^{2}\Gamma^{3}_{0}}{2\pi\omega^{}_{0}}\Bigl [\frac{\mu^{2}_{L}-\mu^{2}_{R}}{(\mu^{2}_{L}+\Gamma^{2}_{0})(\mu^{2}_{R}+\Gamma^{2}_{0})}\Bigr ]^{2}\nonumber\\
&+\frac{\gamma^{2}\Gamma^{3}_{0}}{4\pi}\sum_{\alpha =L,R}\Bigl (\frac{F(\mu^{}_{\alpha},\mu^{}_{\alpha}-\omega^{}_{0})-F(\mu^{}_{\alpha},\mu^{}_{\alpha}+\omega^{}_{0})}{\mu^{2}_{\alpha}+\Gamma^{2}_{0}}+\frac{F(\mu^{}_{\alpha},\mu^{}_{\overline{\alpha}}+\omega^{}_{0})-F(\mu^{}_{\overline{\alpha}},\mu^{}_{\alpha}-\omega^{}_{0})}{\mu^{2}_{\overline{\alpha}}+\Gamma^{2}_{0}}\Bigr )\nonumber\\
&+\frac{\Gamma^{3}_{0}\gamma^{2}\omega^{}_{0}}{4\pi}\ln\Bigl [\frac{(\mu^{}_{L}-\mu^{}_{R}+\omega^{}_{0})^{2}}{\omega^{2}_{0}}\Bigr ]\Bigl (\frac{\Gamma^{2}_{0}-\mu^{}_{L}(\mu^{}_{L}+\omega^{}_{0})}{(\mu^{2}_{L}+\Gamma^{2}_{0})^{2}
((\mu^{}_{L}+\omega^{}_{0})^{2}+\Gamma^{2}_{0})^{2}}+\frac{\Gamma^{2}_{0}-\mu^{}_{R}(\mu^{}_{R}-\omega^{}_{0})}{(\mu^{2}_{R}
+\Gamma^{2}_{0})^{2}
((\mu^{}_{R}-\omega^{}_{0})^{2}+\Gamma^{2}_{0})^{2}} \Bigr )\nonumber\\
&+\frac{\Gamma^{3}_{0}\gamma^{2}\omega^{}_{0}}{4\pi}\ln\Bigl [\frac{(\mu^{}_{L}-\mu^{}_{R}-\omega^{}_{0})^{2}}{\omega^{2}_{0}}\Bigr ]\Bigl (\frac{\Gamma^{2}_{0}-\mu^{}_{L}(\mu^{}_{L}-\omega^{}_{0})}{(\mu^{2}_{L}+\Gamma^{2}_{0})^{2}
((\mu^{}_{L}-\omega^{}_{0})^{2}+\Gamma^{2}_{0})^{2}}+\frac{\Gamma^{2}_{0}-
\mu^{}_{R}(\mu^{}_{R}+\omega^{}_{0})}{(\mu^{2}_{R}
+\Gamma^{2}_{0})^{2}
((\mu^{}_{R}+\omega^{}_{0})^{2}+\Gamma^{2}_{0})^{2}}\Bigr )\ ,\label{GCOS}
\end{align}
\end{widetext}
where $\Delta\mu$ is given by Eq. (\ref{DELTAMU}), and
we have defined
\begin{align}
F(x,y)&=\frac{1}{(y^{2}+\Gamma^{2}_{0})^{2}}\Bigl (\frac{(x+y)(y^{2}+\Gamma^{2}_{0})}{x^{2}+\Gamma^{2}_{0}}\nonumber\\
&-y\ln\Bigl [\frac{x^{2} +\Gamma^{2}_{0}}{\omega^{2}_{0}}\Bigr
]+\frac{\Gamma^{2}_{0}-y^{2}}{\Gamma^{}_{0}}{\rm arctan}
\frac{x}{\Gamma^{}_{0}}\Bigr )\ .
\end{align}
Also, $\overline{\alpha}$ marks the lead which is not $\alpha$.
One observes that  when $\mu_{L}=\mu_{R}$, then $G^{}_{\rm co}$ is
fully given by only the first sum in Eq. (\ref{GCOS}), reducing to
the linear-response result (\ref{Gcolin}).

Finally, the differential conductance resulting from the current
$I^{}_{\rm inco}$, Eq. (\ref{INCOT}), is
\begin{align}
&\frac{2\pi}{e^{2}}G^{}_{\rm inco}=\frac{\gamma^{2}\Gamma^{2}_{0}}{2}
\Theta (\mu^{}_{L}-\mu^{}_{R}-\omega^{}_{0})\nonumber\\
&\times\Bigl (\frac{\mu^{2}_{L}(\mu^{}_{L}-\omega^{}_{0})^{2}-\Gamma^{4}_{0}}{(\mu^{2}_{L}+\Gamma^{2}_{0})_{}^{2}((\mu^{}_{L}-\omega^{}_{0})^{2}+\Gamma^{2}_{0})_{}^{2}}\nonumber\\
&+\frac{\mu^{2}_{R}(\mu^{}_{R}+\omega^{}_{0})^{2}-\Gamma^{4}_{0}}{(\mu^{2}_{R}+\Gamma^{2}_{0})_{}^{2}((\mu^{}_{R}+\omega^{}_{0})^{2}+\Gamma^{2}_{0})_{}^{2}}\Bigr )\ .
\end{align}
At the threshold bias voltage, $eV=\mu^{}_L-\mu^{}_R=\omega^{}_0$,
this contribution jumps from zero to
\begin{align}
&\frac{2\pi}{e^{2}}\Delta G^{}_{\rm
inco}=\gamma^{2}\Gamma^{2}_{0}
\frac{\mu^{2}_L\mu^{2}_R-\Gamma_0^4}{(\mu_L^2+\Gamma_0^2)^2(\mu_R^2+\Gamma_0^2)^2}.
\end{align}
Since at $eV=\omega_{0}$ the chemical potentials are  $\mu_{L}=\mu +\omega_{0}/2$ and $\mu_{R}=\mu -\omega_{0}/2$, it follows that
the conductance jumps {\em downwards}  when the common chemical potential of the leads (measured from the resonance level) is
in the range
\begin{align}
{\rm max}[0,(\omega^{}_{0}/2)^{2}-\Gamma^{2}_{0}]\leq\mu^2\leq
(\omega^{}_{0}/2)^{2}+\Gamma^{2}_{0}\ .\label{CONDCON}
\end{align}
Note that the  range in which the conductance jumps downwards  is
shrinking as $\omega_{0}/\Gamma^{}_{0}$ becomes larger. Since the
``bare" elastic transparency of the junction is given by ${\cal
T}=\Gamma_{0}^{2}/(\mu^{2}+\Gamma^{2}_{0})$, the condition
(\ref{CONDCON}) can be put in the form
\begin{align}
\frac{1}{2+(\omega^{}_{0}/2\Gamma^{}_{0})^{2}}\equiv {\cal T}_1
\leq {\cal T}\leq {\cal T}_2 \equiv {\rm min}[1,
(2\Gamma^{}_0/\omega_0)^2]\ .
\end{align}
One notes that the lower border-line transparency
\cite{paulson,vega} ${\cal T}_1$ reaches the ``universal" value
$1/2$ (with ${\cal T}_2=1$) only \cite{gogolin} in the limit of a
very broad resonance, $\Gamma^{}_0\gg \omega^{}_0$, where the
effects of the vibrational excitations are smeared within the
original resonance. As the ratio $\omega^{}_{0}/\Gamma^{}_{0}$
increases, both ${\cal T}_1$ and ${\cal T}_2$ decrease and approach
each other, so that the region with a negative step in $G^{}_{\rm
inco}$ narrows down in this physically relevant region.

For $V>0$, the logarithmic divergence in $G^{}_{\rm co}$ arises only
due to the last term in Eq. (\ref{GCOS}). Near
$eV=\hbar\omega^{}_0$, the coefficient of this term contains the
factor
$\Gamma_0^2-\mu^{}_L\mu^{}_R=\Gamma_0^2+(\omega^{}_0/2)^2-\mu^2$,
which is positive for all ${\cal T}>{\cal T}_1$. Since the argument
of the log is small near $eV=\hbar\omega^{}_0$, this implies a
negative divergence of this term in this range (and a positive one
for ${\cal T}<{\cal T}_1$). Interestingly, the logarithmic term does
not change sign at ${\cal T}_2$, although the step in $G^{}_{\rm
inco}$ does change sign there. We note that the Kramers-Kronig
relation, relating these two singularities, applies only to the
real and imaginary parts of the self-energy, and not to the corresponding
 contributions to the differential conductance.

\begin{widetext}

\begin{figure}[ht]
\includegraphics[width=5.5cm]{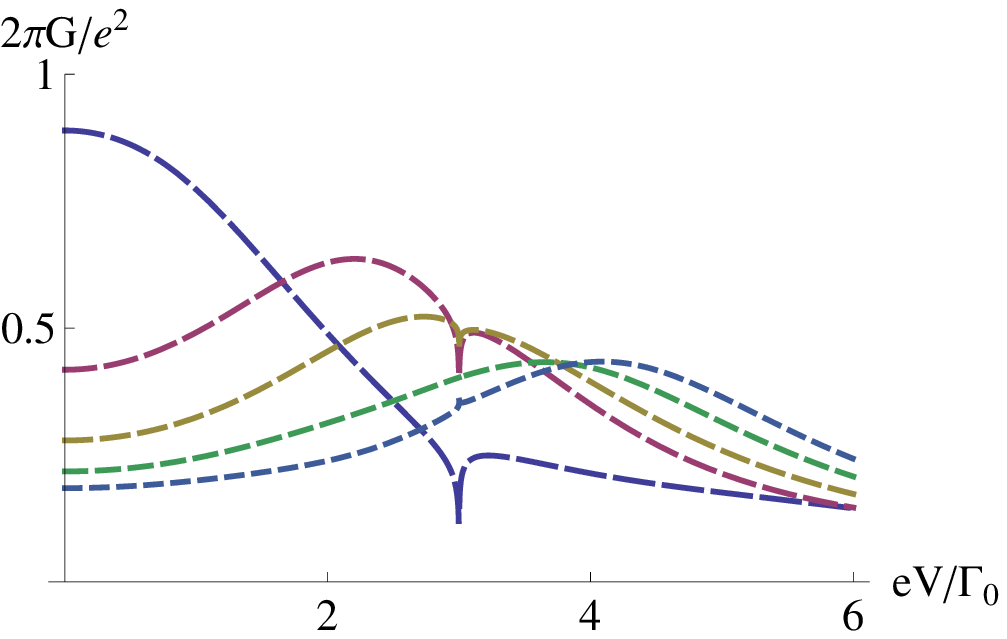}\ \ \
\includegraphics[width=5.5cm]{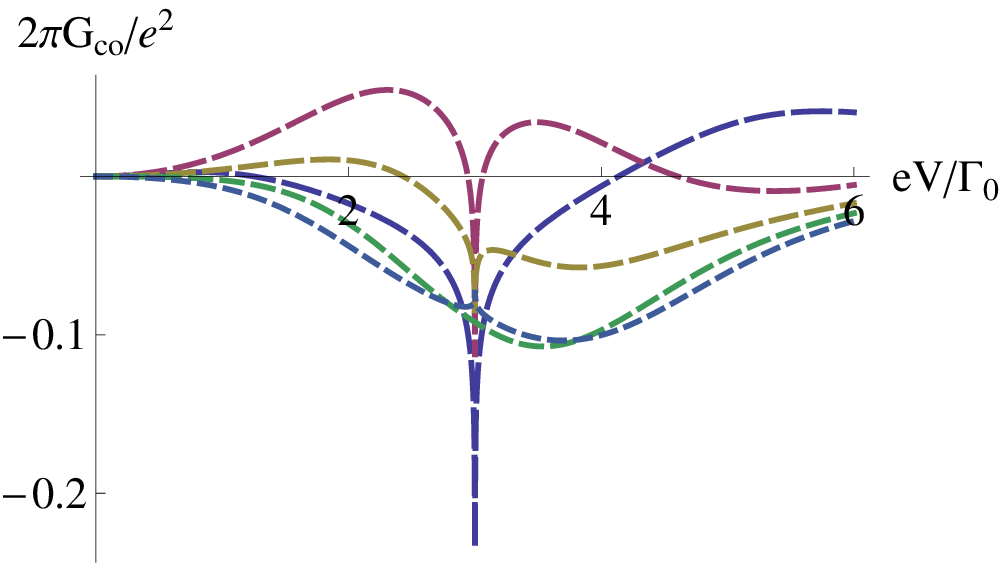}\ \ \
\includegraphics[width=5.5cm]{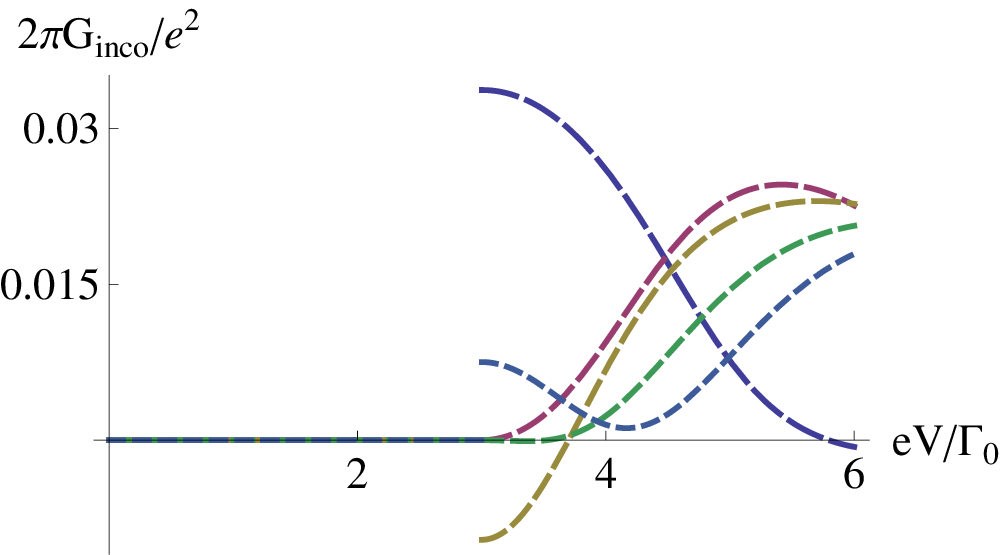}
\caption{The total differential conductance
(left panel)  and the contributions to the conductance from the
coupling to the vibrational mode, $G^{}_{\rm co}$ (middle panel,
containing the logarithmic divergence) and $G^{}_{\rm inco}$ (right
panel, containing the discontinuity; $G^{}_{\rm
inco}=0$ for $eV<\hbar\omega^{}_0$, and the jumps in the various
conductances at the thresholds are seen as the lapses in the
curves)  as function of the bias voltage, for
$\omega^{}_{0}=3\Gamma^{}_{0}$, and five different values of the zeroth-order transmission ${\cal
T}=0.2,~{\cal T}_1=4/17,~0.3,~{\cal T}_2=4/9,~0.9$ (dash sizes increase with ${\cal T}$, appearing in increasing
order at small $V$ in the left panel).  Here $\gamma=\Gamma^{}_{0}$.
}
 \label{om05}
\end{figure}


\end{widetext}

Figure  \ref{om05} shows the total conductance $G=G_{0}+G^{}_{\rm
co}+G^{}_{\rm inco}$, as well as the two separate contributions from
the coupling to the oscillator $G^{}_{\rm co}$ and $G^{}_{\rm inco}$,
for $\omega^{}_0=3\Gamma^{}_0$ and for five values of ${\cal T}$.
The plot for $G^{}_{\rm co}$ does not contain the first term in Eq.
(\ref{GCOS}), which was incorporated into Eq. (\ref{GZEROR}) by the
replacements
$\mu^{}_\alpha\rightarrow\mu^{}_\alpha-\Delta\mu(\mu^{}_\alpha,\mu^{}_L,\mu^{}_R)$.
Clearly, there are no visible singularities when ${\cal T}={\cal
T}_1$, where the coefficients of both the logarithmic term and the
discontinuity vanish (there remain effects for higher derivatives of
the current). The former singularity survives at ${\cal T}={\cal
T}_2$, where the discontinuity vanishes (although more steeply
than near ${\cal T}_1$). The logarithmic divergence is indeed
positive for ${\cal T}<{\cal T}_1$. Also, the magnitudes of both the
jump and the logarithmic divergence are large at large bare
transparencies ${\cal T}$, and decrease with decreasing ${\cal T}$.
It should be kept in mind that the
apparent divergence in
$G^{}_{\rm co}$  results from our expansion in powers of
$\gamma$, which breaks down very close to the threshold
$V=\hbar\omega^{}_0/e$.

\section{Summary}

\label{SUM}

Although several of our formal results have already been obtained
by Mitra {\it et al.} \cite{MITRA} and by Egger and Gogolin
\cite{gogolin}, our paper has used these results for a critical
discussion of several physically relevant issues which have
been debated in the literature. Specifically, we have  obtained and
highlighted the following points:

\noindent (A) At zero temperature, the resonance peak in the
linear-response conductance always narrows down due to the coupling
to the vibrational mode. However, this narrowing down is given by
the Franck-Condon factor only for narrow resonances,  and when one
may ignore the Fermi statistics of the electrons on the leads. When
$\Gamma^{}_0\gg \omega^{}_0$, the electron dwell-time on the dot is
short and therefore the relative narrowing is much smaller.

\noindent (B) Contrary to  claims in the literature, the
linear-response conductance does not exhibit any side-bands at
zero temperature. Small satellites, of order $\exp [-\beta
\omega^{}_0]$, do arise at finite temperatures, where the
excitation of the vibrational mode becomes possible.

\noindent (C) The coupling to the vibrational modes does show up in
the single-particle density of states, which exhibits two
singularities at the frequencies $\omega=\mu\pm\omega^{}_0$ which
correspond to the opening of the inelastic channel in which the
vibrational mode remains excited. (We find only two singular points,
because we expand the results only up to second order in the
coupling to the vibrational mode.) These include discontinuities,
due to the imaginary part of the self-energy, and logarithmic
singularities, due to the real part of the self-energy. The latter
result in deep dips in the density of states around each threshold,
creating apparent side-bands at frequencies which exceed these
thresholds. Although a logarithmic singularity implies the
inapplicability of the perturbative expansion very close to the
inelastic thresholds, the predictions of dips and satellites in the density of states can probably be trusted out of
these narrow regions.

\noindent (D) The same singularities also generate discontinuities
and logarithmic divergences in the differential conductance at and
around the thresholds $eV=\pm \hbar\omega^{}_0$. The signs of the
discontinuities  are usually positive, but they become {\em negative } within a finite range of the
bare elastic transparency of the junction, shrinking progressively
as $\omega^{}_{0}/\Gamma^{}_{0}$ is increased. The ``universal
ratio" \cite{paulson,vega} is obtained only \cite{gogolin} in the
limit $\omega_{0}/\Gamma^{}_{0}\ll 1$, in which the state
with an electron on the dot and the vibrational mode excited is not
well-defined. In contrast, the logarithmic divergences remain
negative for a rather broad range of bare transparencies, indicating
the breakdown of our perturbative expansion very close to the
inelastic thresholds.

\noindent (E) Contrary to some claims in the literature, our results are quite different from those
based on the single-electron transmission, which ignore the Fermi seas in the leads.

It would be useful to test some of these predictions in
experiments or in other theoretical calculations.

\begin{acknowledgments}
This work was initiated by  many illuminating discussions with Doron
Cohen. It was supported by the German Federal Ministry of
Education and Research (BMBF) within the framework of the
German-Israeli project cooperation (DIP), by the US-Israel
Binational Science Foundation (BSF), by the Israel Science
Foundation (ISF) and by the Converging Technologies Program of the
Israel Science Foundation (ISF).
\end{acknowledgments}

\appendix

\section{The Keldysh Green functions}
\label{GREENFK}

We obtain the  Green functions of our system by solving the
Dyson equations up to second order in the coupling $\gamma$. In this procedure, we use the following relation \cite{LANGRETH} for the lesser product of two Green functions,
\begin{align}
(AB)^{<}_{}=A^{r}_{}B^{<}_{}+A^{<}_{}B^{a}_{}\ ,\label{IDEN}
\end{align}
and similarly for the greater product of two Green functions, denoted by the superscript $>$.
Here,
\begin{align}
G^{<}_{ab}(\omega )&=i\int dt e^{i\omega t}\langle b^{\dagger}a(t)\rangle\ ,\nonumber\\
G^{>}_{ab}(\omega )&=-i\int dt e^{i\omega t}\langle a(t)b^{\dagger}\rangle\ .
\label{DEFGK}
\end{align}
Note that when the operators $a$ and $b$ are identical, $G^{<}$ and $G^{>}$ are purely imaginary. Another property of these Green functions (for general $a$ and $b$) is
\begin{align}
G_{}^{<}(\omega )-G^{>}_{}(\omega )=G^{a}_{}(\omega )-G^{r}_{}(\omega )\ ,\label{GKAR}
\end{align}
where
$G^{r}$ ($G^{a}$)  is the retarded (advanced) Green function,
\begin{align}
G^{r}_{ab}(\omega )&=-i\int_{0}^{\infty}e^{i(\omega +i0^{+})t}\langle \Bigl [a(t),b^{\dagger}\Bigr ]^{}_{+}\rangle\ ,\nonumber\\
G^{a}_{ab}(\omega )&=i\int^{0}_{-\infty}e^{i(\omega -i0^{+})t}\langle \Bigl [a(t),b^{\dagger}\Bigr ]^{}_{+}\rangle\ .\label{GRA}
\end{align}
For brevity, the frequency $\omega$ does not appear explicitly in
most of the equations below.

The Dyson equation of the Green function on the dot, $G_{00}$,  reads
\begin{align}
G^{}_{00}&=g^{}_{0}\Bigl (1+\sum_{k}V^{}_{k}G^{}_{k0}+\sum_{p}V^{}_{p}G^{}_{p0}+\gamma G^{}_{0Q0}\Bigr )\ .\label{DG00}
\end{align}
Here, $g_{0}$ is the free Green function of the dot (in the
absence of the coupling with the harmonic oscillator and with the
leads), i.e., $g_{0}=(\omega -\epsilon_{0})^{-1}$. The other Green
functions in Eq. (\ref{DG00}) are those mixing the leads and the
dot operators,
\begin{align}
G^{}_{k(p)0}=\ll c^{}_{k(p)};c^{\dagger}_{0}\gg\ ,\label{GKP0}
\end{align}
and the one mixing the dot and the oscillator operators,
\begin{align}
G^{}_{0Q0}=\ll c^{}_{0}(b+b^{\dagger});c^{\dagger}_{0}\gg\ .\label{OQOQ}
\end{align}
In the notations $\ll;\gg$, the first (second) operator (or a product of operators) is the operator denoted by $a$ ($b$) in Eqs. (\ref{DEFGK}) and (\ref{GRA}).

The Dyson equations of the Green functions (\ref{GKP0})
are
\begin{align}
G^{}_{k(p)0}&=g^{}_{k(p)}V^{}_{k(p)}G^{}_{00}\ .\label{GK0}
\end{align}
Here $g_{k(p)}$ is the free Green function of the left (right) lead,
\begin{align}
g^{r}_{k(p)}&=\frac{1}{\omega -\epsilon^{}_{k(p)}+i0^{+}_{}}=\Bigl (g^{a}_{k(p)}\Bigr )^{\ast}\ ,\nonumber\\
g^{<}_{k(p)}&=(g^{a}_{k(p)}-g^{r}_{k(p)})f^{}_{L(R)}(\omega )\nonumber\\
&=2\pi i\delta (\epsilon^{}_{k(p)}-\omega )f^{}_{L(R)}(\omega )\ ,\label{EXPGLR}
\end{align}
where $f_{L(R)}$ is the Fermi distribution of the left (right) reservoir. As mentioned above, we assume that the two leads are identical except for their different Fermi functions.
It therefore follows that
\begin{align}
\sum_{k}V^{}_{k}G^{}_{k0}+\sum_{p}V^{}_{p}G^{}_{p0}
=\Sigma^{}_{\rm 0}G^{}_{00}\ ,\label{LEADSE}
\end{align}
where
$\Sigma^{}_{0}$ is the self-energy due to the coupling of the dot with the leads,
\begin{align}
\Sigma^{r}_{0}&=\sum_{k}\frac{V^{2}_{k}}{\omega -\epsilon^{}_{k}+i0^{+}}+\sum_{p}\frac{V^{2}_{p}}{\omega -\epsilon^{}_{p}+i0^{+}}\nonumber\\
&\equiv 2\sum_{k}\frac{V^{2}_{k}}{\omega -\epsilon^{}_{k}+i0^{+}}\ ,\label{LSE}
\end{align}
and
\begin{align}
\Sigma^{<}_{0}=&\frac{\Sigma^{a}_{0}-\Sigma^{r}_{0}}{2}\Bigl (f^{}_{R}
+f^{}_{L}\Bigr )\ .\label{XX}
\end{align}
Thus, the Dyson equation (\ref{DG00})  of the dot Green function becomes
\begin{align}
\Bigl (g^{-1}_{0}-\Sigma^{r}_{0}\Bigr )G^{r}_{00}&=1+\gamma G^{r}_{0Q0}\ ,\label{DG00r}
\end{align}
and
\begin{align}
\Bigl (g^{-1}_{0}-\Sigma^{r}_{0}\Bigr )G^{<}_{00}&=
\Sigma^{<}_{0}G^{a}_{00}+\gamma
G^{<}_{0Q0}\ .\label{DG00k}
\end{align}
In particular, the
dot Green functions in the absence of the coupling with the harmonic oscillator, ${\cal G}_{00}$,
are
\begin{align}
{\cal G}^{r}_{00}&=\Bigl (\omega -\epsilon^{}_{0}-\Sigma^{r}_{0}\Bigr )^{-1}\ ,\nonumber\\
{\cal G}^{<}_{00}&={\cal G}^{r}_{00}\Sigma^{<}_{0}{\cal G}^{a}_{00}
=\frac{f^{}_{L}+f^{}_{R}}{2}({\cal G}^{a}_{00}-{\cal G}^{r}_{00})\  .\label{ZERO}
\end{align}

The self-energy coming from the coupling with the harmonic oscillator
results from the Green function   $G_{0Q0}$,  Eq. (\ref{DG00}). Its Dyson equation reads
\begin{align}
G^{}_{0Q0}&=\Bigl (\langle b+b^{\dagger}\rangle \nonumber\\
&+\sum_{k}V^{}_{k}G^{}_{0Qk}+\sum_{p}V^{}_{p}G^{}_{0Qp}+\gamma G^{}_{0Q0Q}\Bigr )g^{}_{0}\ ,\label{0q0}
\end{align}
where
\begin{align}
G^{}_{0Qk(p)}=\ll c^{}_{0}(b+b^{\dagger});c^{\dagger}_{k(p)}\gg\ ,\label{0PK}
\end{align}
and
\begin{align}
G^{}_{0Q0Q}
=\ll c^{}_{0}(b+b^{\dagger});(b+b^{\dagger})c^{\dagger}_{0}\gg\ .\label{G00QQ}
\end{align}
It is straightforward to obtain
\begin{align}
\sum_{k}V^{}_{k}G^{r}_{0Qk}+\sum_{p}V^{}_{p}G^{r}_{0Qp}=\Sigma^{r}_{0}G^{r}_{0Q0}\ ,\label{SUMR1}
\end{align}
and
\begin{align}
\sum_{k}V^{}_{k}G^{<}_{0Qk}+\sum_{p}V^{}_{p}G^{<}_{0Qp}=\Sigma^{a}_{0}
G^{<}_{0Q0}+\Sigma^{<}_{0}G^{r}_{0Q0}\ .\label{SUML1}
\end{align}
Thus we find from Eq. (\ref{0q0}) that
\begin{align}
\Bigl (g^{-1}_{0}-\Sigma^{r}_{0}\Bigr )G^{r}_{0Q0}&=\langle b+b^{\dagger}
\rangle +\gamma G^{r}_{0Q0Q}\ ,\label{DG0Q0r}
\end{align}
and
\begin{align}
\Bigl (g^{-1}_{0}-\Sigma^{a}_{0}\Bigr )G^{<}_{0Q0}&=\Sigma^{<}_{0}G^{r}_{0Q0}
+\gamma G^{<}_{0Q0Q}\ .\label{DG0Q0K}
\end{align}

Inserting the result (\ref{DG0Q0r}) into Eq. (\ref{DG00r}) gives that
the retarded Green function on the dot,  up to second order in the coupling $\gamma$, is
\begin{align}
G^{r}_{00}=\Bigl (\omega -\epsilon^{}_{0}-\gamma\langle b+b^{\dagger}\rangle -\Sigma^{r}_{0}-\Sigma^{r}_{\rm ho}\Bigr )^{-1}\ ,\label{GRSF}
\end{align}
where we have defined
\begin{align}
\Sigma^{r}_{\rm ho}=\gamma^{2}G^{r}_{0Q0Q}\ .\label{SIGHOR}
\end{align}
An analogous result holds for the advanced Green function.
Using this result and Eq. (\ref{DG0Q0K}) in Eq. (\ref{DG00k}) yields
the Keldysh Green function on the dot (again, up to second order in $\gamma$),
\begin{align}
G^{<}_{00}=G^{r}_{00}\Bigl (\Sigma^{<}_{0}+\Sigma^{<}_{\rm ho}\Bigr )G^{a}_{00}\ ,\label{GKSF}
\end{align}
with
\begin{align}
\Sigma^{<}_{\rm ho}=\gamma^{2}G^{<}_{0Q0Q}\ .\label{SIGHO}
\end{align}

It is hence found that the coupling with the harmonic oscillator
modifies the dot Green function in two ways. Firstly, it adds the
term $\Sigma^{}_{\rm ho}=\gamma^{2}G_{0Q0Q}$ to the self-energy.
This contribution is calculated below. Secondly, it shifts the
resonance level by the amount
\begin{align}
\Delta\epsilon^{}_{0}=
\gamma\langle b+b^{\dagger}\rangle =-\frac{2\gamma^{2} }{\omega^{}_{0}}\langle c^{\dagger}_{0}c^{}_{0}\rangle=\frac{2i\gamma^{2} }{\omega^{}_{0}}\int\frac{d\omega}{2\pi}{\cal G}^{<}_{00}(\omega )\ .\label{SHIFT}
\end{align}
This result is found by employing perturbation theory. To first order,
the oscillator wave functions can be written in the form
\begin{align}
\Psi^{}_{n'}=|n'\rangle +\gamma c^{\dagger}_{0}c^{}_{0}\sum_{n\neq
n'}\frac{\langle n|b+b^{\dagger}|n'\rangle}{\omega^{}_{0}(n'
-n)}|n\rangle\ ,\label{PSIN}
\end{align}
and consequently the
diagonal ($n'n'$) matrix element of
$b+b^{\dagger}$ is
\begin{align}
\frac{2\gamma }{\omega^{}_{0}}c^{\dagger}_{0}c^{}_{0}&\sum_{n\neq n'}
\frac{\langle n|b|n'\rangle\langle n' |b^{\dagger}|n\rangle +\langle
n|b^{\dagger}|n' \rangle\langle n' |b^{}|n\rangle}{n' -n}\nonumber\\
&=-
\frac{2\gamma }{\omega^{}_{0}}c^{\dagger}_{0}c^{}_{0}\ .\label{NN}
\end{align}
The average of $c^{\dagger}_{0}c^{}_{0}$
is needed to zeroth order in the coupling with the oscillator,
and therefore is expressed in terms of the Green function (\ref{ZERO}), leading to  Eq. (\ref{SHIFT}).

It remains to compute the Green function $G_{0Q0Q}$, Eq.
(\ref{G00QQ}). As we work up to second order in the coupling
$\gamma$, it is enough to find this function in the absence of the
coupling to the oscillator. At this order, the electron operators
and the oscillator operators are decoupled. For example, using the
definitions of the Keldysh Green functions, Eqs. (\ref{DEFGK}) and
(\ref{GRA}),
\begin{widetext}
\begin{align}
&G^{r}_{0Q0Q}(\omega )=-i\int_{0}^{\infty}dt e^{i\omega t}\Bigl (\langle (b(t)+b_{}^{\dagger}(t))(b+b^{\dagger}_{})\rangle\langle c^{}_{0}(t)c^{\dagger}_{0}\rangle
+\langle (b+b_{}^{\dagger})(b(t)+b^{\dagger}_{}(t))\rangle\langle
c^{\dagger}_{0}c^{}_{0}(t)\rangle \Bigr )\nonumber\\
&=\langle bb^{\dagger}_{}\rangle\int _{0}^{\infty}e^{i(\omega -\omega^{}_{0})t}{\cal G}^{>}_{00}(t)+\langle b^{\dagger}_{}b\rangle
\int _{0}^{\infty}e^{i(\omega +\omega^{}_{0})t}{\cal G}^{>}_{00}(t)-
\langle bb^{\dagger}_{}\rangle\int _{0}^{\infty}e^{i(\omega +\omega^{}_{0})t}{\cal G}^{<}_{00}(t)-\langle b^{\dagger}_{}b\rangle
\int _{0}^{\infty}e^{i(\omega -\omega^{}_{0})t}{\cal G}^{<}_{00}(t)\nonumber\\
&=\int\frac{d\omega '}{2\pi}e^{-i\omega 't}\Bigl (\langle bb^{\dagger}_{}\rangle\int _{0}^{\infty}e^{i(\omega -\omega^{}_{0})t}{\cal G}^{>}_{00}(\omega ')
+\langle b^{\dagger}_{}b\rangle
\int _{0}^{\infty}e^{i(\omega +\omega^{}_{0})t}{\cal G}^{>}_{00}(\omega ')\nonumber\\
&-
\langle bb^{\dagger}_{}\rangle\int _{0}^{\infty}e^{i(\omega +
\omega^{}_{0})t}{\cal G}^{<}_{00}(\omega ')
-\langle b^{\dagger}_{}b\rangle
\int _{0}^{\infty}e^{i(\omega -\omega^{}_{0})t}{\cal G}^{<}_{00}(\omega ')\Bigr )\ .
\end{align}
Therefore, upon carrying out the time-integrations,  we obtain
\begin{align}
G^{r(a)}_{0Q0Q}(\omega )
&=i\int\frac{d\omega '}{2\pi}\Bigl \{ {\cal G}^{>}_{00}(\omega ')\Bigl  (\frac{\langle bb^{\dagger}\rangle}{\omega -\omega^{}_{0}-\omega '\pm i 0^{+}}
+\frac{\langle b^{\dagger}b^{}\rangle}{\omega +\omega^{}_{0}-\omega '\pm i0^{+}}\Bigr )\nonumber\\
&-{\cal G}^{<}_{00}(\omega ')\Bigl  (\frac{\langle bb^{\dagger}\rangle}{\omega +\omega^{}_{0}-\omega '\pm i0^{+}}
+\frac{\langle b^{\dagger}b^{}\rangle}{\omega -\omega^{}_{0}-\omega '\pm i0^{+}}\Bigr )\Bigr \}\ .\label{SIGHO1}
\end{align}
Similarly,
\begin{align}
&G^{<}_{0Q0Q}(\omega )=i\int dt e^{i\omega t}\langle (b+b_{}^{\dagger})(b(t) +b^{\dagger}_{}(t))\rangle \langle c^{\dagger}_{0} c^{}_{0}(t)\rangle \nonumber\\
&=i\langle bb^{\dagger}_{}\rangle \int dt e^{i(\omega +\omega^{}_{0})t}
\rangle \langle c^{\dagger}_{0} c^{}_{0}(t)\rangle +
i\langle b^{\dagger}_{} b\rangle \int dt e^{i(\omega -\omega^{}_{0})t}
\rangle \langle c^{\dagger}_{0} c^{}_{0}(t)\rangle \ ,\label{GRHO}
\end{align}
\end{widetext}
and consequently
\begin{align}
&G^{<}_{0Q0Q}(\omega )=
\langle b^{\dagger}b\rangle {\cal G}^{<}_{00}(\omega -\omega^{}_{0})+\langle bb^{\dagger}\rangle {\cal G}^{<}_{00}(\omega +\omega^{}_{0})\ ,\nonumber\\
&G^{>}_{0Q0Q}(\omega )=\langle b^{\dagger}b\rangle {\cal G}^{>}_{00}(\omega +\omega^{}_{0})+\langle bb^{\dagger}\rangle {\cal G}^{>}_{00}(\omega -\omega^{}_{0})\ ,\label{GSOFG}
\end{align}
where ${\cal G}_{00}$ is the Green function on the dot in the absence of the coupling with the oscillator, see Eqs. (\ref{ZERO}).
It is easy to check that at zero temperature and {\em at equilibrium}, Eq. (\ref{GRHO})
reduces to the usual diagrammatic expression, see e.g., Ref.~ \onlinecite{LEVY}.

In order to present explicit expressions for the self-energy due to the harmonic oscillator, we use  [see Eqs.  (\ref{XX}) and (\ref{ZERO})]
\begin{align}
{\cal G}^{<}_{00}(\omega )&=i|{\cal G}^{r}_{00}(\omega )|^{2}\Gamma^{}_{0}(\omega )\Bigl (f^{}_{L}(\omega )+f^{}_{R}(\omega )
\Bigr )\ ,\nonumber\\
{\cal G}^{>}_{00}(\omega )&=i|{\cal G}^{r}_{00}(\omega )|^{2}\Gamma^{}_{0}(\omega )\Bigl (f^{}_{L}(\omega )+f^{}_{R}(\omega )
-2\Bigr )\ ,
\end{align}
where we have denoted
\begin{align}
\Gamma^{}_{0}(\omega )=\frac{\Sigma^{a}_{0}(\omega )-\Sigma^{r}_{0}(\omega )}{2i}\ .
\label{GAM0}
\end{align}
It follows that [see Eqs.  (\ref{SIGHOR}) and (\ref{SIGHO1})]
\begin{widetext}
\begin{align}
\Sigma^{\stackrel{r}{a}}_{\rm ho}(\omega )
=\gamma^{2}\int\frac{d\omega '}{2\pi}|{\cal G}^{r}_{00}(\omega ' )|^{2}_{}\Gamma^{}_{0}(\omega  ')&\Biggl  (\frac{\langle bb^{\dagger}\rangle (2-f^{}_{L}(\omega ')-f^{}_{R}(\omega '))+\langle b^{\dagger}_{}b\rangle (f^{}_{L}(\omega ')+f^{}_{R}(\omega '))}{\omega -\omega^{}_{0}-\omega '\pm i 0^{+}}\nonumber\\
&
+\frac{\langle b^{\dagger}b^{}\rangle(2-f^{}_{L}(\omega ')-f^{}_{R}(\omega '))+\langle bb^{\dagger}_{}\rangle (f^{}_{L}(\omega ')+f^{}_{R}(\omega '))}{\omega +\omega^{}_{0}-\omega '\pm i0^{+}}\Biggr )\ .\label{SIGHORsof}
\end{align}
\end{widetext}
It is instructive to interpret Eq. (\ref{SIGHORsof}) in the simple
equilibrium case where $f_{L}(\omega ')=f_{R}(\omega ')=f(\omega ')$
as the change, within second-order perturbation theory in $\gamma$,
of the energy of an electronic state at energy $\omega$, due to all
other states, at a running energy $\omega '$. The term $|{\cal
G}^{r}_{00}(\omega ' )|^{2}_{}\Gamma^{}_{0}(\omega ')$ appearing
before the large brackets is just the density of the latter states
at zero-order in $\gamma$. For the imaginary part of
$\Sigma^{r(a)}_{\rm ho}(\omega )$, the first term in the large
brackets is due to real transitions occurring by exciting the
oscillator (intensity proportional to $\langle bb^{\dagger}\rangle
$) and going to $\omega '=\omega -\omega^{}_{0}$ with the blocking
factor $1-f(\omega ')$, or by absorbing a `phonon' (intensity
proportional to $\langle b^{\dagger}b\rangle $) and going to
$\omega$ from the same $\omega '$, now with an initial population
$f(\omega ')$. The real part of $\Sigma^{r(a)}_{\rm ho}(\omega )$,
given by the principal part of the integrals, is just the
corresponding perturbation-theory energy shift. Obviously, these
real and imaginary parts satisfy the Kramers-Kronig relationships.
The second term in the large brackets is likewise understood as
involving transitions to the state $\omega '=\omega +\omega^{}_{0}$
(for the imaginary part) or to the states around it (for the real
part).


\begin{thebibliography}{999}
\bibitem{COMMENT}

In Ref. ~ \onlinecite{gogolin}, the part of the current arising
from the real part of $\Sigma^{}_{\rm ho}$ is termed
`quasielastic', while the one involving the imaginary part is
called `inelastic'.
\end{thebibliography}
\end{document}